\documentclass{article}

\usepackage{arxiv}

\usepackage[utf8]{inputenc} % allow utf-8 input
\usepackage[T1]{fontenc}  % use 8-bit T1 fonts

\usepackage{url}      % simple URL typesetting
\usepackage{booktabs}    % professional-quality tables
\usepackage{amsfonts}    % blackboard math symbols
\usepackage{nicefrac}    % compact symbols for 1/2, etc.
\usepackage{microtype}   % microtypography
\usepackage{lipsum}

\usepackage{threeparttable}
\usepackage{amsmath}
\usepackage{hyperref}
\usepackage{cleveref}

\usepackage{graphicx}
\usepackage{adjustbox}
\usepackage{rotating} 
\usepackage{multirow}
\usepackage{caption}
\usepackage{blindtext}
\usepackage{tabularx,ragged2e}
\usepackage[toc,page]{appendix}
\usepackage{tablefootnote}

\newcommand{\de}{\mathrm{d}}

\newcommand{\bY}{\mathbf{Y}}
\newcommand{\bd}{\mathbf{D}}

\newcolumntype{C}{>{\Centering\arraybackslash}X} % centered "X" column

\newtheorem{theorem}{Theorem}[section]

\newtheorem{lemma}[theorem]{Lemma}
\title{Overall Agreement for Multiple Raters with Replicated Measurements}

\author{
 Tongrong Wang \\
 Biostatistics\&Bioinformatices\\
 School of Medicine, Duke University\\
 Durham, North Carolina, U.S.A. \\
 \texttt{ tongrong.wang@duke.edu} \\
  \And
 Huiman X. Barnhart\\
 Biostatistics\&Bioinformatices\\
 School of Medicine, Duke University\\
 Durham, North Carolina, U.S.A. \\
 \texttt{huiman.barnhart@duke.edu} \\
}

\begin{document}
\maketitle

\begin{abstract}
Multiple raters are often needed to be used interchangeably in practice for measurement or evaluation. Assessing agreement among these multiple raters via agreement indices are necessary before their participation. While the intuitively appealing agreement indices such as coverage probability and total deviation index, and relative area under coverage probability curve, have been extended for assessing overall agreement among multiple raters, these extensions have limitations. The existing overall agreement indices either require normality and homogeneity assumptions or did not preserve the intuitive interpretation of the indices originally defined for two raters. In this paper, we propose a new set of overall agreement indices based on maximum pairwise differences among all raters. The proposed new overall coverage probability, overall total deviation index and relative area under overall coverage probability curve retain the original intuitive interpretation from the pairwise version. Without making any distributional assumption, we also propose a new unified nonparametric estimation and inference approach for the overall indices based on generalized estimating equations that can accommodate replications made by the same rater. Under mild assumptions, the proposed variance estimator is shown to achieve efficiency bound under independent working correlation matrix. Simulation studies under different scenarios are conducted to assess the performance of the proposed estimation and inference approach with and without replications. We illustrate the methodology by using a blood pressure data with three raters who made three replications on each subjects.
\end{abstract}

% keywords can be removed
\keywords{Overall Agreement; coverage probability; total deviation index; relative area under the coverage probability curve; replication; generalized estimating equations}

\section{Introduction}
Indices are developed to assess the agreement among different raters or different measurements from a rater on the same subjects. Barnhart et al. categorized existing approaches for evaluation of agreement into descriptive tools, scaled summary indices attaining values between -1 and 1 and unscaled summary indices\cite{barnhart2007overview}. She further elucidated that unscaled indices, such as coverage probability (CP) and total deviation index(TDI), are preferred for assessing the agreement in a core lab setting with following advantages: (1) they are simple to implement ; (2) they can be interpreted intuitively in terms of the original measurement unit; (3) the CP can provide actionable results that guide readers to identify the source of the agreement to improve their measurements\cite{barnhart2016choice}. CP quantifies the chance of the difference between two measurements from two given raters on the same subject being less than a pre-fixed acceptable boundary $\delta$. TDI, as a counterpart of CP, is the boundary where the difference of two measurement falls into with a pre-specified confidence or probability. In practice, it may be desirable to set more than one acceptable/tolerable boundary up to a maximum acceptable difference $\delta_{max}$ with corresponding acceptable CPs. For example, the British hypertension society protocol (BHSP) for the evaluation of blood pressure measuring device \cite{o1993british} shown in Table \ref{tab::bhsp_bp}. This protocol classifies the grade of blood pressure measuring devices by specify the satisfactory CPs for multiple pre-specified differences (Table \ref{tab::bhsp_bp}). Therefore, it is useful to summarize the agreement based on a coverage probability curve defined as the curve of coverage probability for a range of differences. A relative area under coverage probability curve (RAUCPC) was introduced as a summary index as a measure of agreement\cite{raucpc}.
\begin{table}[ht!]
\centering
\caption{The British hypertension society protocol for the evaluation of blood pressure measuring device}
\label{tab::bhsp_bp}
\begin{tabular}{ccccc}
\hline
\multicolumn{5}{c}{Pre-specified Difference (mmHg)}  \\ \hline
Grade & $\leq 5$ & $\leq 10$ & $\leq 15$ & $\leq 20$ \\ \hline
\multicolumn{5}{c}{Pre-specified Coverage Probability} \\ \hline
A   & 60    & 85    & 95    & 100    \\
B   & 50    & 75    & 90    & 95    \\
C   & 40    & 65    & 85    & 90    \\
D   & \multicolumn{4}{c}{Fail to achieve C}     \\ \hline
\end{tabular}\\
    {\small 20mmHg is not included in the original protocol but added here as the maximum acceptable difference \par}

\end{table}

CP, TDI and RAUCPC are all originally defined for two raters. However, there may be some competing new raters developed at the same time that need to be compared to each other or to an existing rater. Here we are interested in the interchangeability among more than two raters. For example, in the data set published in the Bland and Altman's paper \cite{bland1986}, the blood pressures of 85 patients were measured by two human observers and one device. We are interested in whether these three raters (two human observers and one device) can be used interchangeably. Therefore, it is desirable to extend the CP, TDI and RAUCPC to measure agreement among multiple raters while preserving the intuitive interpretation of pairwise version. Lin et al.\cite{lin2007unified} first extended the concept of CP and TDI to multiple raters using two-way mixed model and later Jang et al.\cite{jang2018overall} proposed a new set of definitions based on the root mean square of pairwise differences(RMSPD). Although these overall agreement indices give some insight of the closeness of tested raters, they have limitations due to assumptions that may not hold in practice. First, the overall unscaled indices proposed by Lin is an approximate measure that are good only when following assumptions hold\cite{lin2007unified}: (1) the relative bias square is small; (2) the measurements follow normal distributions; (3) homogeneity across different raters. For overall indices proposed by Jang et al.\cite{jang2018overall}, though the assumptions are relaxed, it still requires the difference from two measurements on the same subject is normally distributed. We will demonstrate these assumptions do not hold for the blood pressure measurements\cite{bland1986} in section \ref{sec::BP}. 

In addition to the distributional assumption, the overall indices proposed by Jang et al.\cite{jang2018overall} is difficult to have intuitive interpretation in practice. For example, for the overall CP(OCP) proposed by Jang et al., the satisfactory boundary is specified based on the root mean square of all pairwise differences among the raters. While an acceptable difference between two measurements can be chosen based on clinical implication, but it is not easy to choose an acceptable root mean square of differences in practice because its magnitude is difficult to interpret in terms of clinical judgement. Moreover, a satisfactory agreement through this OCP cannot guarantee the raters are interchangeabe. For example, if one rater has large departure from the rest raters and thus not interchangeable with others, but by averaging squared pairwise differences the resulting RMSPD can be acceptable and leading to claiming all raters are interchangeable. Similar problems could happen for their overall TDI and RAUCPC as well. Last but not the least, Jang's method\cite{jang2018overall} cannot be applied to the measurements by raters with replications and would need to be applied by restricting to one measurement per rater on each subject.

To address aforementioned issues, we propose new sets of overall CP, TDI and RAUCPC based on the maximum pairwise distances (MPD) to assess overall agreement among all considered raters. The new indices have intuitive interpretation in terms of original measurement unit and can be estimated by a unified non-parametric distribution-free paradigm based on Generalized Estimation Equations (GEE). The GEE approach could assess the inter- and intra-rater agreement simultaneously without normality and homogeneity assumptions. Moreover, under minimum set of assumptions, we show that the the estimator will achieve the semi-parametric efficiency bound using the working independence covariance matrix. The paper is organized as follow. In Section 2, we first propose a new estimator of pairwise RAUCPC. Then we introduce the new definitions of overall CP, TDI and RAUCPC and with the new estimator of RAUCPC, we are able to develop a unified GEE approach for estimation and inference for all overall indices. We provide the simulation results in assessing the performance of the unified approach and illustrate the method with the example of the blood pressure data from Bland and Altman\cite{bland1986} in Section 3 and Section 4 respectively. Finally, we draw conclusions and provide some discussions in Section 5.

\section{Methods}
We are interested in whether $J$ raters can be used interchangeably for making same type of measurement in a given population. For a subject randomly sampled from a population, we denote $Y_{j}$ as the measurement taken by rater $j, j=1, ..., J$. The interchangeability among the raters can be based on a distance metric, $D$, which reflects the closeness among measurements given by the raters. For $J=2$, the distance metrics is defined as 
\begin{align}
D=|Y_{1}-Y_{2}|
\end{align}
It is intuitive that a smaller distance implies a better agreement between two raters. For using these two raters interchangeably, one would like to have a high probability that this distance is within an acceptable difference. Therefore, the concept of coverage probability is used and it is defined as the probability that the distance falls within a boundary $d$,
\begin{align}
  CP(d)=Pr(D<d)=F_D(d)
\end{align}
where $F_D(d)$ is the cumulative distribution function of $D$ over the target population. The higher the CP(d), the better the agreement. To use CP to claim satisfactory agreement, we need to pre-defined a clinically acceptable boundary $\delta_0$ and the corresponding satisfactory probability $\pi_0$. If $CP(\delta_0)$ is greater than or equal to $\pi_0$, we can claim that two raters are interchangeable. The TDI, as a counterpart of CP, is defined as the boundary that $\pi$ proportion of the distances falls within
\begin{align}
TDI(\pi)=F^{-1}_D(\pi)
\end{align}
The smaller the TDI is, the better the agreement is. To use TDI to claim satisfactory agreement, we need to pre-defined a clinically acceptable probability $\pi_0$ and a satisfactory boundary $\delta_0$. If $TDI(\pi_0)$ is less than or equal to $\delta_0$, then we claim that two raters are interchangeable. 

\begin{figure}[ht!]
  \centering
  \includegraphics[scale=0.5]{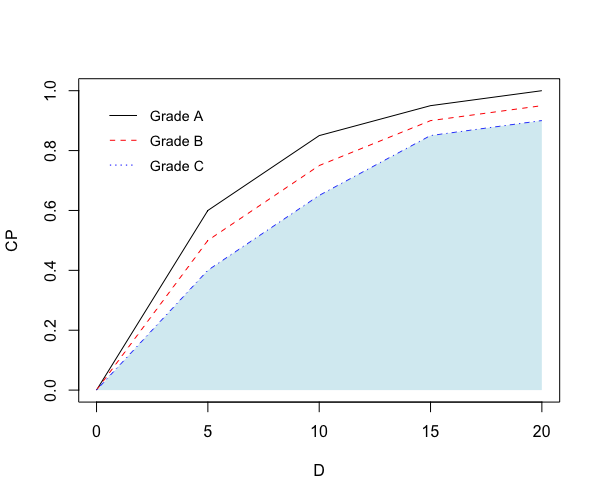}
  \caption{Agreement criteria based on RAUCPC for three different grades of BP devices.}
  \label{fig::bhsp}
\end{figure}

In practice, it is sometimes desirable to control the quality on more than one pre-specified differences, or we may simply want to summarize the agreement based on differences up to a maximum acceptable difference. This leads to consider a coverage probability curve $(d, CP(d))$. Relative area under the CP curve (RAUCPC) is proposed \cite{barnhart2016choice} as a summary index of coverage probability curve by the scaled area under the coverage probability curve to a maximum acceptable difference $\delta_{max}$. The $\delta_{max}$ is often chosen as $CP(\delta_{max})=1$. Specifically, RAUCPC is defined as 
\begin{align}
  RAUCPC(\delta_{max})=\frac{\int_{0}^{\delta_{max}}CP(d)\de d}{\delta_{max}}=\frac{\int_{0}^{\delta_{max}}F_D(d)\de d}{\delta_{max}}
\end{align}
RAUCPC ranges from 0 to 1 and a greater value indicates a better agreement. To use RAUCPC to claim satisfactory agreement, we need to pre-specify an acceptable RAUCPC, $\tau_0$. If RAUCPC is greater than or equal to $\tau_0$, the we claim that two raters are interchangeable. It is not obvious how to choose $\tau_0$ and we can use the British hypertension society protocol to illustrate one way of choosing $\tau_0$. As shown is Table \ref{tab::bhsp_bp}, we can set $\delta_{max}=20$. For grade C device, satisfactory coverage probabilities are specified for the absolute differences of 5, 10, 15, and 20. By linearly connecting these specific points, $(d, CP(d))$ for $d=5,10,15,20$, yields a satisfactory coverage probability curve for grade C device. Curves corresponding to Grade A, B and C are shown in Figure \ref{fig::bhsp}. The shaded area for Grade C is equal to 11.8 and the the RAUCPC is 0.59. Thus, one can use $\tau_0=0.59$ as the criterion for satisfactory Grade C device. Similar $\tau_0$ can be computed for claiming Grade A or Grade B device. 

Jang et al.\cite{jang2018overall} extended the pairwise CP, TDI and RAUCPC to multiple raters by defined the distance metrics as
\begin{align}
D^*=\sqrt{\frac{2}{J(J-1)}\sum_{1\leq p<q\leq J}(Y_p-Y_q)^2}
\end{align}
Their overall CP based on the $D^*$ was defined as
$OCP^*(d)=Pr(D^*<d)$
However, this definition may claim a satisfactory overall agreement and fail to detect the outlier raters due to the averaging. For example, suppose there are four competing raters where the first three raters give identical measurements and the fourth rater gives measurements greater than the other three raters by 5 on all subjects. Suppose the pre-defined clinical meaningful boundary is $\delta_{CP}=3$. Then $D^*=2.8$ for all subjects implying $\hat{OCP^*}(3)=1$ and thus it would satisfactory agreement, even through the fourth rater gave clinically different outcomes on every subjects with the other three raters. Thus there is a need for a better distance metrics for assessing agreement among multiple raters 

\subsection{Proposed Overall Agreement Indices}

For two raters, aforementioned agreement indices defined interchangeability in the sense that whichever rater measures the subject the results are clinically similar. We want to extend this intuition of interchangeability for the situation with more than two raters. As a example, a patient walks into a local clinic and has his/her blood pressure measured by one of the multiple nurses there. The patient would expect that his/her blood pressure measurement will be similar no matter which nurse takes the measurement, in the sense that the nurses are interchangeable. Therefore, the largest difference between any two nurses should be clinically negligible.
Following this idea, we define a new distance metric, maximum pairwise difference (MPD), among $J$ raters as
\begin{align}
  D=\max\lbrace |Y_{j}-Y_{j'}|: j,j'=1,\cdots,J \rbrace \label{equ::defmpd}
\end{align}
The maximum pairwise difference could avoid the pitfalls brought by averaging all pairwise difference and can be reduced to the pairwise difference when $J=2$. The new overall coverage probability (OCP) and overall totally deviation index (OTDI) based on MPD are defined as
\begin{align}
OCP(d)=Pr(D\leq d)=F_{D}(d) \qquad OTDI(\pi)=F^{-1}_{D}(\pi)
\end{align}
where $F_{D}(d)$ is the cumulative distribution function of MPD. The OCP measures the probability that maximum pairwise difference among $J$ raters are less than a given acceptable boundary $\delta_{0}$ with clinical meaning. To use OCP to claim satisfactory agreement, we need to pre-define a clinically acceptable boundary $\delta_0$ and the corresponding satisfactory probability $\pi_0$. Since the PMD is in the same unit of pairwise distance, $\delta_0$ can be chosen similarly with the pairwise CP. If $OCP(\delta_0) \ge \pi_0$, then chance that the measurements given by J raters on the same subject are within distance $\delta_0$ is greater than $\pi_0$.  
As a counterpart of OCP, OTDI captures the boundary that all possible pairwise differences among the raters of $100\pi_0$\% subjects are fall into (-OTDI($\pi_0$), OTDI($\pi_0$)) and if OTDI($\pi_0$) is smaller than the pre-set satisfactory boundary $\delta_0$ then we could claim that the raters are interchangeable. 

 When there are more than one acceptable boundary is of interest or we are interested in an aggregated agreement up to a maximum boundary $\delta_{max}$, the overall relative area under the overall coverage probability curve, as an extension of RAUCPC, is defined as
\begin{align}
  RAUOCPC(\delta_{max})=\frac{\int_{0}^{\delta_{max}}OCP(d)\de d}{\delta_{max}}=\frac{\int_{0}^{\delta_{max}}F_{D}(d)\de d}{\delta_{max}}
\end{align}
Like RAUCPC, RAUOCPC ranges from 0 to 1 and higher values indicates a good agreement. Since the MPD preserves the original unit of the measurement, we can use the clinical information from the pairwise version about the acceptable boundary for setting multiple boundaries for MPD. In this way, we could set the satisfactory RAUOCPC the same way as we did for the pairwise RAUCPC.

\subsection{Estimation and Inference}
We propose a unified distribution-free GEE approach to estimate and make inference on OCP, OTDI and RAUOCPC. Minimum assumptions are made as follows: (1) measurements of different subject are independent; (2) replicated measurements are i.i.d. given the same rater on the same subject. Since current estimators for RAUCPC proposed by Barnhart\cite{raucpc} can not be expressed as a sum of function of each subjects, they are unable to fit in the GEE framework. Therefore, in this section, in order to have a unifed GEE model for RAUOCPC together with OCP and OTDI, we first propose a new unbiased non-parametric estimator for pairwise RAUCPC and RAUOCPC. Second, we present the unified GEE model and the inference approach.

\subsubsection{Unbiased Non-parametric Estimator for RAUCPC}
Barnhart\cite{raucpc} proposed both parametric and non-parametric approaches for estimating RAUCPC. The parametric RAUCPC estimator is calculated based on the estimated density function of $D$ where the measurements are assumed to follow a normal distribution. While for the non-parametric estimator, suppose all distinct observations of $D$ are $d_1<\cdots<d_m$ with the corresponding estimated values of $\hat{CP}(d_i)$. Let $d_0=0$ and $d_{m+1}=\delta_{max}$ , the estimated RAUCPC is equal to the area under the connected line $(d_i,\hat{CP}(d_i))$ based on trapesoid rule. It is clear that both estimators cannot be expressed as a sum of independent functions of individual subjects. Therefore, a new unbiased nonparametric estimator is developed below for our unified GEE framework. 

The RAUCPC is defined as the scaled integration of cumulative distribution function of distance metric from 0 to a maximum acceptable boundary $\delta_{max}$. By the role of integration by parts,
\begin{align}
  \int_{0}^{\delta_{max}}F_D(d)\de d &= \left. F_D(d)d \right|^{\delta_{max}}_0 - \int_{0}^{\delta_{max}}d f_D(d) \de d\\
  &=\delta_{max}\int_{0}^{\delta_{max}}f_D(d)\de d -\int_{0}^{\delta_{max}}d f_D(d) \de d\\
  &=\int_{0}^{\delta_{max}}(\delta_{max}- d)f_D(d)\de d \label{equ::raucpc_exp}
\end{align}
We note that equation (\ref{equ::raucpc_exp}) has the form of expectation for the following new random variable,
\begin{align}
C&=
\begin{cases}
\delta_{max}-D & 0\leq D <\delta_{max}\\
0  & D \geq \delta_{max}
\end{cases}\\
&=\max(0,\delta_{max}-D )
\end{align}
Then equation (\ref{equ::raucpc_exp}) can be expressed as $E[C]$ which is proofed in the Appendix \ref{app::raucpc}. This leads to the following lemma.

 \begin{lemma}\label{lemma::raucpc}
 The relative area under coverage probability curve can be expressed as, i.e.,
 \begin{align}
  RAUCPC=& \frac{\int_{0}^{\delta_{max}}F_D(d)\de d}{\delta_{max}}=\frac{E[C]}{\delta_{max}}
\end{align}
 \end{lemma}
Similarly, for RAUOCPC with multiple raters, $RAUOCPC=E[C]/ \delta_{max}$.
Followed by the Lemma \ref{lemma::raucpc}, we can use moment estimator $\frac{1}{N}\sum_{i=1}^N C_i/\delta_{max}$ for RAUCPC/RAUOCPC when there is no replications. This form of estimator can be easily incorporated into GEE framework when there are replicates.

The performance of this new non-parametric estimator of RAUCPC is assessed via simulation and the results are presented in Appendix \ref{app::simu_newraucpc}. In general, for both normally and non-normally distributed data, the new proposed non-parametric estimator has similar or better performance than the previous estimators\cite{raucpc} in terms of bias and mean square error(MSE).

\subsubsection{Unified Generalized Estimation Equation Approach}
 Let $Y_{ijk}$ be the $k$th successive replicate measured by $j$th rater on $i$th subject. It is reasonable to assume that successive replicates measured by same rater on same subject are equivalent where $\{Y_{ij1},\cdots,Y_{ijK}\}$, are identically and independent distributed when we condition on $i$th subject and $j$th rater. This assumption implies that the unconditional distribution of $\mathbf{Y}_{ij}=(Y_{ij1},\cdots,Y_{ijK})$ has a distribution with an exchangeable correlation matrix $\Sigma_j$ where we denote $cor(Y_{ijk},Y_{ij'k'})=\rho_{jj'} \forall k, k'$ , $cor(\bY_{ij},\bY_{ij'})=\rho_{jj'} \mathbf{1}_{K_j\times K_{j'}}$ , $\mathbf{1}_{K\times K}$ is a $K_{j} \times K_{j'}$ matrix with 1 as elements and $K_j$ is the number of replicates for $j${th} rater. For simplicity, we assume that the number of replicates are equal to $K$ for all raters per subjects and it can be easily extended to unbalanced design. Since MPD is defined over a collection of J measurements with one from each rater on the same subject, when there are $K$ replications of each raters, we can get $K^J$ distinct collections of $\{Y_{i1k_{1}},\cdots,Y_{iJk_{J}}\}$, $k_j=1,\cdots,K$. On each collection, we could compute the observed MPD and index it by $m$. The MPD at the $m$th collection, $\{k_1^m,\cdots,k_J^m\}$, is expressed as
\begin{align}
  D_{im}=max\lbrace|Y_{ijk_j^m}-Y_{ij'k_{j'}^m}|:j,j'=1,\cdots,J\rbrace, \quad m=1,\cdots, K^J \label{equ::tcor}
\end{align}
If $J=2$, then this distance reduces to the distance for two raters mentioned above. For a random subject $i$, we have a random vector $\bd_i=(D_{i1},\cdots, D_{iM})$ where $D_{im}$ has the same marginal distribution $F_D(d)$. 

To develop a unified form, we first denote the agreement index of interest, OCP, OTDI or RAUOCPC, as $\beta$ and channel it with the parameter for estimation, $\theta$, in GEE model with link function $\theta=g(\beta)$. For OCP and RAUOCPC ranging from 0 to 1, logit transformation is used. For OTDI, since it is greater than 0, the natural log transformation is used. After transformation, the parameter of interest, $\theta$, in GEE model can range form $-\infty$ to $\infty$. Under the standard GEE framework\cite{liang1986longitudinal}, we need to find a function $s(D_{im},\beta)$ such that $\mathrm{E}\left[s(D_{im},\beta)\right]=0$. Let $I(\cdot)$ be the indicator function, we choose the following $s(\cdot)$ corresponding to different agreement parameters of interest
\begin{align}
 &s(D_{im},\theta) =I(D_{i1}<\delta_{0})-g(\theta) &\beta=OCP(\delta_0)=g(\theta)=\frac{\exp(\theta)}{1+\exp(\theta)} \\
&s(D_{im},\theta) =\pi_{0}-I(D_{i1}-g(\theta)<0) &\beta=OTDI(\pi_0)=g(\theta)=\exp(\theta) \\
&s(D_{im},\beta) =\frac{\max(0,\delta_{max}-D_{i1})}{\delta_{max}}-g(\theta) &\beta=RAUOCPC(\delta_{max})=g(\theta)=\frac{\exp(\theta)}{1+\exp(\theta)} 
 \end{align}
Now we construct the generalized estimating equations system based on $\{\bd_i\}_{i=1,\cdots,N}$ as follow,
 \begin{align}
 \sum_{i=1}^{N} \Gamma_i^T V_{i}^{-1}S_i=0 \label{equ::gee0}
 \end{align}
 $\mathbf{S}_i=\left(s(D_{i1},\theta),\cdots,s(D_{iM},\theta)\right)^T$, $\Gamma_i=\frac{\partial S_i}{\partial \theta}$, $V_i=A_i^{1/2}R_w(\alpha)A_i^{1/2}$ with working correlation matrix $R_w(\alpha)$ and $A_i=diag(\Sigma_S)$ where $\Sigma_S$ is the covariance matrix of $S_i$. For OCP and RAUOCPC, we have $\frac{\partial s(D_{im},\theta)}{\partial \theta}=\frac{\exp(\theta)}{(1+\exp(\theta))^2}$ which does not depend on $D_{im}$. 
 
 For OTDI, we note that $s(D_{im},\theta)$ is not differentiable with regard to $\theta$ and a different definition of $\Gamma_i$ in equation (19) is needed. Rather than differentiating $s(D_{im},\theta)$, we would differentiate its expectation, $\pi_0-F_D(D_{im}$, where $F_D(d)$ is the marginal cumulative distribution function of $D_{im}$. Then, $\Gamma_i=1_{n}f_D(\exp(\theta))\exp(\theta)$ where $f_D(\exp(\theta))$ is the marginal density function of $D_{im}$ at point $\exp(\theta)$. The marginal distribution, $f_D(\theta)$, is a nuisance parameter and needs to be estimated in order to solve equation (19).A smoothed kernel density\cite{duong2019package} can be used and implemented by using a R function {\it kde} in R 3.1.1.  With this new $\Gamma_i$ in (19), we show that the limiting distribution of $\hat{\theta}$ still follows the general results of GEE in Appendix \ref{app::OTDI}.
 
 Therefore, across three kinds of agreement indices, $\Gamma_i$ can be expressed as 
\begin{align}
\Gamma_i=\mathbf{1}_{M\times 1}h(\theta)\label{equ::gamma}
\end{align}
where $h(\theta)=\frac{\exp(\theta)}{(1+\exp(\theta))^2}$ for OCP and RAUOCPC and $h(\theta)=f_D(\exp(\theta))\exp(\theta)$ for OTDI.

Now we consider the specification of working correlation matrix $R_w(\alpha)$ with nuisance parameter $\alpha$. The optimal asymptotic efficiency of $\hat{\theta}$ is achieved when $R_w(\alpha)$ coincides with the true correlation matrix of $D_i$\cite{wang2005effects}. A misspecified working correlation matrix could greatly compromised the efficiency especially when sample size is small and design is unbalance\cite{wang2003working}. Therefore, generally a working correlation matrix resembling the truth is desirable. However, for our model, using the independent working correlation matrix will obtain the same efficiency (see theorem below) as the true correlation matrix. This may seem to be surprising, but it is due to the unique structure of true correlation matrix for agreement data as shown in the lemma below. Therefore, we will use the independent working correlation matrix in equation (\ref{equ::gee0}) without the need to estimate the nuisance parameter $\alpha$ in the working correlation matrix. 

\begin{lemma}
\label{lemma::euqalrow}
Given $Y_{ijk}|i,j$ are i.i.d, the sum of elements in each row of correlation matrix of $S_i$, $R_0$, are equal.
\end{lemma}
The proof of lemma \ref{lemma::euqalrow} is in Appendix \ref{app::eqeualrow}. Based on this lemma, we have the following theorem on efficiency. 
 \begin{theorem}\label{th::eq0}
The GEE estimator obtained from equation (\ref{equ::gee0}) will achieve the same asymptotically statistical efficiency under either the true correlation matrix or the independent working correlation matrix. The limiting distribution of $\hat{\theta}$ is
\begin{align}
n^{-1/2}(\hat{\theta}-\theta_0)\rightarrow \mathcal{N}\left(0,\frac{a\sigma^2_S}{Mh^2(\theta_0)}\right)\label{equ::limitd}
\end{align}

where $a$ is the row sum of $R_0$, $\sigma_S^2=var(s(D_{im},\theta))$ and $M=K^J$. 
\end{theorem}

The proof is in Appendix \ref{app:geeq}. The variance form in equation (\ref{equ::limitd}) is the simplification of the robust sandwich estimator after utilizing equation (\ref{equ::gamma}) and Lemma \ref{lemma::euqalrow}. The estimation from equation (\ref{equ::gee0}) and inference  (\ref{equ::limitd}) can be obtained by standard statistical software that implements GEE. Moreover, the results can be easily extended to the unbalance design with $M=\prod_{j=1}^JK_j $ where $K_j$ is the number of replicates for $j$th rater.

Our main interest is to determine whether the considered raters can be used interchangeably. This can be determined by performing a hypothesis test on one of the three indices depending on the nature of the question. The hypothesis can be formed as one of the following, 
\begin{align}
H_0: OCP(\delta_0) <\pi_0 \quad &versus \quad H_a: OCP(\delta_0) \geq \pi_0 \label{equ::hcp} \\
H_0: OTDI(\delta_0) >\delta_0 \quad &versus \quad H_a: OTDI (\delta_0) \leq \delta_0  \label{equ::htdi}\\
H_0: RAUOCPC(\delta_{max}) <\tau_0 \quad &verseus \quad H_a: RAUOCPC(\delta_{max}) \geq\tau_0 \label{equ::hr}
\end{align}
If the null hypothesis is rejected, we can claim that the considered raters are interchangeable. Moreover, note here we use one-sided test instead of two-sided test, since the primary interest is to determine if the raters can be used interchangeably. For example, suppose the satisfactory OCP is 0.9 for an acceptable difference $\delta_0$ which means we will claim the raters are interchangeable if more than 90\% of the MPD is within $\delta_0$. One would not want to frame the hypothesis as $H_0: OCP(\delta_0)=0.9 \quad \text{versus} \quad H_a: OCP(\delta_0)\neq 0.9$, because we would reject the null hypothesis either with low or high OCP, e.g., when $OCP(\delta_0)=0.5$ or $OCP(\delta_0)=0.95$. It would not make sense to claim the interchangeability among the raters when $OCP(\delta_0)=0.5$ by rejecting the null. Therefore, hypotheses like (\ref{equ::hcp}), (\ref{equ::htdi}) or (\ref{equ::hr}) make more sense for agreement studies.

\section{Simulation}
We assess the performance of the new proposed overall indices by using simulated data from both normal and log-normal distributions. Suppose each subject is measured by three raters with $J=3$. Let $Y_{ij}=(Y_{ij1},\cdots,Y_{ijK})$ be the $K$ replicates measured by rater $j$ on subject $i$ and $Y_i=(Y_{i1},Y_{i2},Y_{i3})^T$ be $3K\times 1$ vector of measurements of subject $i$. Without loss of generosity, we simulated data so that the mean and covariance matrix of $Y_i$ have the following forms
\begin{align}
\mathbf{\mu}&= (\mu_{1} \mathbf{1}_{1\times K},\mu_{2} \mathbf{1}_{1\times K},\mu_{3} \mathbf{1}_{1\times K})^T \label{equ::simus1}\\
\Sigma_Y&=\begin{pmatrix}
\sigma_1^2 \Sigma_1 & \sigma_1 \sigma_2\rho_{12} \mathbf{1}_{K\times K} & \sigma_1 \sigma_3\rho_{13} \mathbf{1}_{K\times K} \\
 \sigma_1 \sigma_2 \rho_{12} \mathbf{1}_{K\times K} & \sigma_2^2 \Sigma_2 & \sigma_2 \sigma_3\rho_{23} \mathbf{1}_{K\times K} \\
 \sigma_1 \sigma_3\rho_{13} \mathbf{1}_{K\times K} & \sigma_2 \sigma_3\rho_{23} \mathbf{1}_{K\times K} & \sigma_3^2 \Sigma_3
\end{pmatrix}\label{equ::simus2}\\
\Sigma_j &=\begin{pmatrix}
1 & \rho_j & \rho_j\\
\rho_j & 1 & \rho_j \\
\rho_j &\rho_j &1
\end{pmatrix} \quad \text{for j=1,2 and 3}\label{equ::simus3}
\end{align}
where $\sigma^2_i$ is intra-rater variance, $\rho_{jj'}$ represent the correlation between rater $j$ and $j'$ and  $\rho_j$. For log-normal data $Y_i$ is generated by taking exponential transformation of a random vector from a multivariate normal distribution specified in Appendix \ref{app::lognormal} so that the resulting $Y_i$ has the above form of mean and covariance matrix. 

 The agreement indices and the correlation matrix of $D_i$ from equation (\ref{equ::tcor}) cannot be expressed allegorically as functions of the mean and covariance matrix parameters for the specified normal and log-normal distributions. Therefore, numerical approximation is used to obtain the true values of the agreement indices and correlation matrix of $D_i$ given the true parameters of the normal and log-normal distributions. Specifically one simulated data set with a huge sample size of 100,000 is generated to represent the true population. True agreement indices are obtained by using the corresponding GEE estimators of the agreement indices based on this large data set and the true correlation matrix of $D_i$ is obtained by the observed sample correlation matrix of the observed $D_i$. 
 
 we set acceptable difference of $\delta_{0}=4$ and $\delta_{0}=3.5$ for normal and log-normal data respectively in OCP, acceptable probability of $\tau_0=0.8$ in OTDI and maximum acceptable difference of $\delta_{max}=5$ and $\delta_{max}=4$ for normal and log-normal data respectively in RAUOCPC. To set the parameters for the normal and log-normal distributions, we leverage the intra- and inter-rater correlation and systematic shift between raters to achieve the following four different agreement scenarios and the resulting true values of  OCP, OTDI and RAUOCPC are shown in table \ref{tab::simuout_cp}-\ref{tab::simuout_r} for each scenarios.
 
\begin{itemize}
\item High agreement: no systematic shift in means $\mathbf{\mu}=(1,1,1)$ and  high correlation $\rho_{j}=0.8$ and $\rho_{jj'}=0.5$ for $j,j'=1,2,3$
\item Moderate agreement: no systematic shift in means $\mathbf{\mu}=(1,1,1)$ and low correlation $\rho_{j}=0.5$ and $\rho_{jj'}=0.1$ for $j,j'=1,2,3$
\item Mild agreement: systematic shift in means $\mathbf{\mu}=(3,1,1)$ and high correlation $\rho_{j}=0.8$ and $\rho_{jj'}=0.5$ for $j,j'=1,2,3$
\item Low agreement: systematic shift in means $\mathbf{\mu}=(3,1,1)$ and  low correlation $\rho_{j}=0.5$ and $\rho_{jj'}=0.1$ for $j,j'=1,2,3$ 
\end{itemize}
For all four parameter scenarios, we set the intra-rater variability be  $\sigma^2_1=\sigma^2_1=2$ and $\sigma_3^2=1$ to represent some heterogeneity across the raters. Designs without replicates and with replicates of 3 are simulated and we consider sample size of 20, 30, 100, and 500. Together with the four different parameter settings, this resulted in a total of 48 simulation scenarios.  For each scenario, a total of 10,000 simulated data sets are generated. 

The performance of the proposed GEE approach are evaluated by reporting bias of estimated agreement indexes, mean square error(MSE), standard deviation(SD) of the 10,000 estimated agreement indexes and coverage rate (CR), where CR is defined as the percentage of estimated one-sided 95\% confident interval cover the true value. Moreover, the standard errors of estimators are calculated using the true correlation matrix, $se_t$ , and independent working correlation matrix, $se_{ind}$ , to confirm our theoretical result in Theorem \ref{th::eq0}.  The results are shown in table \ref{tab::simuout_cp}, table \ref{tab::simuout_tdi} and table \ref{tab::simuout_r} for OCP, OTDI and RAUOCPC respectively. 

In general, the simulation results show that the bias is negligible for both normal and non-normal data sets even when the sample size is as small as 20, since the proposed approach is unbiased and does not rely on the normality and homogeneity assumption. The results from data sets with replicates outperform the those for data without replicates in terms of bias and MSE. When sample size is small, the CR is closer to $95\%(?)$ for data with the replicates than those without replicates. Moreover, for all three indices and different sample sizes, the absolute difference between the robust sandwich estimator with independent working correlation matrix $se_{ind}$ and the one with true correlation matrix $se_{t}$ is less than or equal than 0.001 which confirms Theory \ref{th::eq0}. 

For OCP, as shown in Table \ref{tab::simuout_cp}, the OCP varies from 50\% to over 90\% for different combination of correlation and mean values. For data sets without replicates, some OCP estimates are unidentifiable when the true OCP exceeds 80\% and sample size is under 30, since all PMDs are smaller than the pre-determined satisfactory boundary $\delta_{0}$. A reasonable CR around 94\% can be achieved for such data sets for sample size of 100 or larger. While for the data sets with replicates, all OCP estimations are well defined and the 94.3\% CR can be achieved for sample size of 20. As shown in Table \ref{tab::simuout_tdi}, the true TDI varies from 1.3825 to 4.0512 for different combination of correlation and mean values. When sample size is small, the CR is over 96\% for data sets without replicates which may due to the inaccuracy of estimating kernel function with limited sample. While a reasonable CR around 92\% to 96\% can be achieved for the data sets with replicates across all sample sizes. For RAUOCPC, we set $\delta_{max}=5$ and $\delta_{max}=4$ for normal and log-normal data respectively. As shown in Table \ref{tab::simuout_r}, the true RAUOCPC varies from 0.3101 to 0.7719 for different combination of correlation and mean values. The RAUOCPC does not encounter the same problems when sample size is small as the OCP and OTDI. The CR is between 94\% and 96\% for all parameter scenarios.

\begin{sidewaystable}

\centering
\begin{threeparttable}

\caption{Simulation Results of Overall Coverage Probability}
\label{tab::simuout_cp}
  
\begin{tabular}{rrccrrrrrrrrrrrrrrrrr}
\hline
\multicolumn{1}{c}{\multirow{2}{*}{nsub}} & \multicolumn{1}{c}{\multirow{2}{*}{nrep}} & \multirow{2}{*}{Corr} & \multirow{2}{*}{Shift} & \multicolumn{8}{c}{Normal} & \multicolumn{1}{c}{} & \multicolumn{8}{c}{Lognormal} \\ \cline{5-12} \cline{14-21} 
&&&&\multicolumn{1}{c}{True}&&&&&&&&&\multicolumn{1}{c}{True}\\
\multicolumn{1}{c}{} & \multicolumn{1}{c}{} &  &  & \multicolumn{1}{c}{OCP} & \multicolumn{1}{c}{Bias} & \multicolumn{1}{c}{$SD_{t}$\tnote{a}} & \multicolumn{1}{c}{$SE_{ind}$\tnote{b}} & \multicolumn{1}{c}{$SE_{t}$\tnote{c}} & \multicolumn{1}{c}{MSE} & \multicolumn{1}{c}{CR} & \multicolumn{1}{c}{nmiss \tnote{d}} & \multicolumn{1}{c}{} & \multicolumn{1}{c}{OCP} & \multicolumn{1}{c}{Bias} & \multicolumn{1}{c}{$SD_{t}$\tnote{a}} & \multicolumn{1}{c}{$SE_{ind}$\tnote{b}} & \multicolumn{1}{c}{$SE_{t}$\tnote{c}} & \multicolumn{1}{c}{MSE} & \multicolumn{1}{c}{CR} & \multicolumn{1}{c}{nmiss\tnote{d}} \\ \hline
20 & 1 & high & No & 0.9412 & 0.0008 & 0.4976 & 0.8738 &   & 0.0027 & 100.0\% & 3000 &  & 0.9444 & 0.0007 & 0.4949 & 0.8787 &   & 0.0026 & 100.0\% & 3284 \\
20 & 3 & high & No & 0.9412 & -0.0003 & 0.6777 & 0.5875 & 0.5730 & 0.0010 & 90.4\% & 0 &  & 0.9444 & -0.0054 & 0.7850 & 0.7597 & 0.7418 & 0.0015 & 95.8\% & 0 \\
50 & 1 & high & No & 0.9412 & -0.0003 & 0.5869 & 0.6520 &   & 0.0011 & 92.3\% & 464 &  & 0.9444 & 0.0001 & 0.5826 & 0.6673 &   & 0.0011 & 93.8\% & 576 \\
50 & 3 & high & No & 0.9412 & -0.0001 & 0.4071 & 0.3731 & 0.3696 & 0.0004 & 93.0\% & 0 &  & 0.9444 & -0.0001 & 0.5821 & 0.5071 & 0.5031 & 0.0007 & 93.5\% & 0 \\
100 & 1 & high & No & 0.9412 & 0.0000 & 0.4770 & 0.4553 &   & 0.0006 & 92.8\% & 24 &  & 0.9444 & 0.0002 & 0.4860 & 0.4695 &   & 0.0005 & 94.9\% & 39 \\
100 & 3 & high & No & 0.9412 & 0.0000 & 0.2724 & 0.2634 & 0.2623 & 0.0002 & 94.0\% & 0 &  & 0.9444 & -0.0003 & 0.3653 & 0.3489 & 0.3479 & 0.0003 & 95.2\% & 0 \\
500 & 1 & high & No & 0.9412 & 0.0001 & 0.1953 & 0.1925 &   & 0.0001 & 95.4\% & 0 &  & 0.9444 & 0.0001 & 0.1988 & 0.1977 &   & 0.0001 & 95.1\% & 0 \\
500 & 3 & high & No & 0.9412 & -0.0001 & 0.1169 & 0.1177 & 0.1176 & 0.0000 & 95.3\% & 0 &  & 0.9444 & -0.0001 & 0.1547 & 0.1534 & 0.1536 & 0.0001 & 95.1\% & 0 \\ \hline
20 & 1 & low & No & 0.8066 & -0.0013 & 0.6119 & 0.6091 &   & 0.0080 & 98.5\% & 147 &  & 0.8970 & 0.0003 & 0.6022 & 0.7696 &   & 0.0046 & 100.0\% & 1154 \\
20 & 3 & low & No & 0.8066 & -0.0004 & 0.3386 & 0.3279 & 0.3201 & 0.0026 & 93.9\% & 0 &  & 0.8970 & -0.0003 & 0.5979 & 0.5398 & 0.5267 & 0.0022 & 93.3\% & 0 \\
50 & 1 & low & No & 0.8066 & -0.0005 & 0.3855 & 0.3696 &   & 0.0031 & 95.1\% & 0 &  & 0.8970 & -0.0001 & 0.5236 & 0.5011 &   & 0.0019 & 93.0\% & 33 \\
50 & 3 & low & No & 0.8066 & -0.0001 & 0.2080 & 0.2065 & 0.2048 & 0.0010 & 94.4\% & 0 &  & 0.8970 & -0.0001 & 0.3481 & 0.3316 & 0.3286 & 0.0009 & 94.6\% & 0 \\
100 & 1 & low & No & 0.8066 & -0.0004 & 0.2601 & 0.2568 &   & 0.0016 & 93.7\% & 0 &  & 0.8970 & 0.0002 & 0.3509 & 0.3409 &   & 0.0009 & 95.4\% & 0 \\
100 & 3 & low & No & 0.8066 & 0.0000 & 0.1469 & 0.1459 & 0.1454 & 0.0005 & 94.7\% & 0 &  & 0.8970 & -0.0002 & 0.2357 & 0.2320 & 0.2311 & 0.0005 & 95.2\% & 0 \\
500 & 1 & low & No & 0.8066 & 0.0000 & 0.1147 & 0.1136 &   & 0.0003 & 95.0\% & 0 &  & 0.8970 & 0.0000 & 0.1473 & 0.1480 &   & 0.0002 & 94.6\% & 0 \\
500 & 3 & low & No & 0.8066 & 0.0000 & 0.0651 & 0.0651 & 0.0651 & 0.0001 & 95.0\% & 0 &  & 0.8970 & 0.0000 & 0.1029 & 0.1030 & 0.1030 & 0.0001 & 95.0\% & 0 \\ \hline
20 & 1 & high & Yes & 0.6457 & -0.0015 & 0.4974 & 0.4845 &   & 0.0112 & 96.3\% & 3 &  & 0.7898 & 0.0007 & 0.5991 & 0.5904 &   & 0.0083 & 99.0\% & 96 \\
20 & 3 & high & Yes & 0.6457 & -0.0005 & 0.3494 & 0.3480 & 0.3395 & 0.0060 & 95.0\% & 0 &  & 0.7898 & 0.0000 & 0.4748 & 0.4452 & 0.4344 & 0.0051 & 94.9\% & 0 \\
50 & 1 & high & Yes & 0.6457 & -0.0001 & 0.3040 & 0.2997 &   & 0.0046 & 95.2\% & 0 &  & 0.7898 & 0.0000 & 0.3649 & 0.3570 &   & 0.0033 & 95.4\% & 1 \\
50 & 3 & high & Yes & 0.6457 & -0.0005 & 0.2171 & 0.2170 & 0.2150 & 0.0024 & 95.0\% & 0 &  & 0.7898 & -0.0004 & 0.2794 & 0.2744 & 0.2719 & 0.0020 & 95.1\% & 0 \\
100 & 1 & high & Yes & 0.6457 & -0.0004 & 0.2102 & 0.2104 &   & 0.0023 & 95.3\% & 0 &  & 0.7898 & -0.0001 & 0.2509 & 0.2487 &   & 0.0016 & 94.3\% & 0 \\
100 & 3 & high & Yes & 0.6457 & -0.0003 & 0.1522 & 0.1527 & 0.1521 & 0.0012 & 95.2\% & 0 &  & 0.7898 & -0.0005 & 0.1935 & 0.1925 & 0.1917 & 0.0010 & 95.1\% & 0 \\
500 & 1 & high & Yes & 0.6457 & 0.0000 & 0.0924 & 0.0936 &   & 0.0004 & 94.7\% & 0 &  & 0.7898 & 0.0000 & 0.1105 & 0.1100 &   & 0.0003 & 95.7\% & 0 \\
500 & 3 & high & Yes & 0.6457 & 0.0000 & 0.0675 & 0.0681 & 0.0681 & 0.0002 & 95.3\% & 0 &  & 0.7898 & 0.0000 & 0.0852 & 0.0856 & 0.0856 & 0.0002 & 95.0\% & 0 \\ \hline
20 & 1 & low & Yes & 0.5397 & 0.0011 & 0.4706 & 0.4615 &   & 0.0122 & 95.5\% & 0 &  & 0.6802 & -0.0012 & 0.5230 & 0.5004 &   & 0.0108 & 97.5\% & 6 \\
20 & 3 & low & Yes & 0.5397 & -0.0012 & 0.2922 & 0.2935 & 0.2863 & 0.0050 & 95.4\% & 0 &  & 0.6802 & 0.0001 & 0.3632 & 0.3532 & 0.3445 & 0.0057 & 94.6\% & 0 \\
50 & 1 & low & Yes & 0.5397 & -0.0005 & 0.2903 & 0.2868 &   & 0.0050 & 93.6\% & 0 &  & 0.6802 & -0.0002 & 0.3094 & 0.3078 &   & 0.0043 & 95.3\% & 0 \\
50 & 3 & low & Yes & 0.5397 & -0.0004 & 0.1843 & 0.1841 & 0.1825 & 0.0021 & 95.2\% & 0 &  & 0.6802 & -0.0004 & 0.2235 & 0.2204 & 0.2183 & 0.0023 & 95.1\% & 0 \\
100 & 1 & low & Yes & 0.5397 & -0.0003 & 0.2022 & 0.2017 &   & 0.0025 & 95.3\% & 0 &  & 0.6802 & 0.0002 & 0.2186 & 0.2161 &   & 0.0022 & 94.6\% & 0 \\
100 & 3 & low & Yes & 0.5397 & -0.0003 & 0.1298 & 0.1297 & 0.1292 & 0.0010 & 94.7\% & 0 &  & 0.6802 & -0.0005 & 0.1545 & 0.1550 & 0.1543 & 0.0011 & 95.2\% & 0 \\
500 & 1 & low & Yes & 0.5397 & -0.0001 & 0.0899 & 0.0898 &   & 0.0005 & 94.9\% & 0 &  & 0.6802 & -0.0001 & 0.0958 & 0.0960 &   & 0.0004 & 95.2\% & 0 \\
500 & 3 & low & Yes & 0.5397 & 0.0001 & 0.0573 & 0.0579 & 0.0579 & 0.0002 & 95.3\% & 0 &  & 0.6802 & 0.0001 & 0.0685 & 0.0691 & 0.0691 & 0.0002 & 95.2\% & 0 \\ \hline
\end{tabular}

\begin{tablenotes}\footnotesize
\item [a] Standard deviation of 10,000 estimated OCPs which should be close to the true standard error of the estimator
\item [b] Mean estimated standard error of estimators from GEE with independent correlation matrix
\item [c] Mean estimated standard error of estimators from GEE with true correlation matrix (This value is left blank when there is no replicates because it is the same as $SE_{ind}$)
\item [d] Number of simulations with estimated OCP equalling 100\% which lends to undefined value in the logit function and the results are based on the outcomes without this issue
\end{tablenotes}
  \end{threeparttable}
\end{sidewaystable}

\begin{sidewaystable}
\footnotesize
\centering
\begin{threeparttable}
\caption{Simulation Results of Overall Total Deviation Index}
\label{tab::simuout_tdi}
  \begin{tabular}{rrccrrrrrrrrrrrrrrr}
\hline
\multicolumn{1}{c}{\multirow{2}{*}{nsub}} & \multicolumn{1}{c}{\multirow{2}{*}{nrep}} & \multirow{2}{*}{correlation} & \multirow{2}{*}{Shift} & \multicolumn{7}{c}{Normal} & \multicolumn{1}{c}{} & \multicolumn{7}{c}{Lognormal} \\ \cline{5-11} \cline{13-19} 
&&&&\multicolumn{1}{c}{True}&&&&&&&&\multicolumn{1}{c}{True}\\
\multicolumn{1}{c}{} & \multicolumn{1}{c}{} &  &  & \multicolumn{1}{c}{OTDI} & \multicolumn{1}{c}{Bias} & \multicolumn{1}{c}{$SD_t$\tnote{a}} & \multicolumn{1}{c}{$SE_{ind}$\tnote{b}} & \multicolumn{1}{c}{$SE_{t}$\tnote{c}} & \multicolumn{1}{c}{MSE} & \multicolumn{1}{c}{CR} & \multicolumn{1}{c}{} & \multicolumn{1}{c}{OTDI} & \multicolumn{1}{c}{Bias} & \multicolumn{1}{c}{$SD_t$\tnote{a}} & \multicolumn{1}{c}{$SE_{ind}$\tnote{b}} & \multicolumn{1}{c}{$SE_{t}$\tnote{c}} & \multicolumn{1}{c}{MSE} & \multicolumn{1}{c}{CR} \\ \hline
20   & 1    & High        & No    & 2.2455 & -0.0852 & 0.1380 & 0.1657          &               & 0.0948 & 91.6\% &  & 1.3808 & -0.0658 & 0.3113 & 0.3401          &               & 0.1853 & 87.8\%  \\
20   & 3    & High        & No    & 2.2455 & -0.0003 & 0.0896 & 0.0938          & 0.0938        & 0.0405 & 93.3\% &  & 1.3808 & 0.0690  & 0.2554 & 0.2670          & 0.2671        & 0.1547 & 93.2\%  \\
50   & 1    & High        & No    & 2.2455 & -0.0328 & 0.0901 & 0.1004          &               & 0.0406 & 93.0\% &  & 1.3808 & -0.0222 & 0.2044 & 0.2134          &               & 0.0811 & 90.7\%  \\
50   & 3    & High        & No    & 2.2455 & 0.0014  & 0.0587 & 0.0595          & 0.0595        & 0.0173 & 93.7\% &  & 1.3808 & 0.0137  & 0.1673 & 0.1654          & 0.1655        & 0.0562 & 92.1\%  \\
100  & 1    & High        & No    & 2.2455 & -0.0155 & 0.0626 & 0.0693          &               & 0.0197 & 94.2\% &  & 1.3808 & -0.0126 & 0.1468 & 0.1497          &               & 0.0411 & 91.7\%  \\
100  & 3    & High        & No    & 2.2455 & 0.0016  & 0.0407 & 0.0419          & 0.0419        & 0.0084 & 94.6\% &  & 1.3808 & 0.0093  & 0.1185 & 0.1173          & 0.1173        & 0.0278 & 93.2\%  \\
500  & 1    & High        & No    & 2.2455 & -0.0020 & 0.0283 & 0.0297          &               & 0.0040 & 95.2\% &  & 1.3808 & -0.0017 & 0.0651 & 0.0662          &               & 0.0081 & 93.8\%  \\
500  & 3    & High        & No    & 2.2455 & 0.0020  & 0.0182 & 0.0186          & 0.0186        & 0.0017 & 95.4\% &  & 1.3808 & 0.0017  & 0.0523 & 0.0526          & 0.0526        & 0.0052 & 94.6\%  \\ 
\hline
20   & 1    & Low         & No    & 2.9660 & -0.1058 & 0.1404 & 0.1677          &               & 0.1702 & 91.4\% &  & 2.0300 & -0.0935 & 0.2820 & 0.3068          &               & 0.3296 & 87.8\%  \\
20   & 3    & Low         & No    & 2.9660 & -0.0016 & 0.0813 & 0.0838          & 0.0839        & 0.0581 & 92.8\% &  & 2.0300 & 0.0272  & 0.2060 & 0.2069          & 0.2072        & 0.1883 & 90.9\%  \\
50   & 1    & Low         & No    & 2.9660 & -0.0430 & 0.0896 & 0.1008          &               & 0.0703 & 92.5\% &  & 2.0300 & -0.0390 & 0.1820 & 0.1923          &               & 0.1374 & 91.1\%  \\
50   & 3    & Low         & No    & 2.9660 & 0.0002  & 0.0512 & 0.0529          & 0.0530        & 0.0230 & 94.1\% &  & 2.0300 & 0.0119  & 0.1340 & 0.1316          & 0.1317        & 0.0769 & 92.3\%  \\
100  & 1    & Low         & No    & 2.9660 & -0.0201 & 0.0628 & 0.0696          &               & 0.0346 & 94.1\% &  & 2.0300 & -0.0198 & 0.1305 & 0.1351          &               & 0.0703 & 91.9\%  \\
100  & 3    & Low         & No    & 2.9660 & 0.0010  & 0.0366 & 0.0373          & 0.0373        & 0.0118 & 94.1\% &  & 2.0300 & 0.0090  & 0.0932 & 0.0933          & 0.0934        & 0.0366 & 93.7\%  \\
500  & 1    & Low         & No    & 2.9660 & -0.0024 & 0.0285 & 0.0299          &               & 0.0071 & 95.0\% &  & 2.0300 & -0.0024 & 0.0588 & 0.0595          &               & 0.0142 & 93.7\%  \\
500  & 3    & Low         & No    & 2.9660 & 0.0024  & 0.0163 & 0.0165          & 0.0166        & 0.0024 & 95.2\% &  & 2.0300 & 0.0024  & 0.0412 & 0.0417          & 0.0418        & 0.0070 & 94.7\%  \\ 
\hline
20   & 1    & High        & Yes   & 3.5219 & -0.0995 & 0.1009 & 0.1224          &               & 0.1275 & 92.2\% &  & 3.0327 & -0.0703 & 0.0973 & 0.1059          &               & 0.0904 & 87.9\%  \\
20   & 3    & High        & Yes   & 3.5219 & -0.0089 & 0.0726 & 0.0750          & 0.0751        & 0.0646 & 92.7\% &  & 3.0327 & 0.0029  & 0.0807 & 0.0785          & 0.0785        & 0.0613 & 89.9\%  \\
50   & 1    & High        & Yes   & 3.5219 & -0.0395 & 0.0639 & 0.0734          &               & 0.0508 & 93.3\% &  & 3.0327 & -0.0292 & 0.0630 & 0.0660          &               & 0.0371 & 90.6\%  \\
50   & 3    & High        & Yes   & 3.5219 & -0.0006 & 0.0459 & 0.0474          & 0.0474        & 0.0260 & 94.1\% &  & 3.0327 & 0.0034  & 0.0506 & 0.0500          & 0.0501        & 0.0238 & 92.2\%  \\
100  & 1    & High        & Yes   & 3.5219 & -0.0195 & 0.0454 & 0.0503          &               & 0.0257 & 94.2\% &  & 3.0327 & -0.0133 & 0.0455 & 0.0463          &               & 0.0192 & 91.7\%  \\
100  & 3    & High        & Yes   & 3.5219 & 0.0004  & 0.0327 & 0.0333          & 0.0333        & 0.0133 & 94.5\% &  & 3.0327 & 0.0034  & 0.0355 & 0.0355          & 0.0355        & 0.0116 & 93.3\%  \\
500  & 1    & High        & Yes   & 3.5219 & -0.0020 & 0.0204 & 0.0216          &               & 0.0052 & 95.2\% &  & 3.0327 & -0.0016 & 0.0202 & 0.0205          &               & 0.0038 & 93.8\%  \\
500  & 3    & High        & Yes   & 3.5219 & 0.0020  & 0.0145 & 0.0148          & 0.0148        & 0.0026 & 95.3\% &  & 3.0327 & 0.0016  & 0.0157 & 0.0159          & 0.0159        & 0.0023 & 94.7\%  \\ 
\hline
20   & 1    & Low         & Yes   & 4.0533 & -0.1252 & 0.1152 & 0.1385          &               & 0.2171 & 92.0\% &  & 3.4661 & -0.0987 & 0.1201 & 0.1317          &               & 0.1786 & 88.0\%  \\
20   & 3    & Low         & Yes   & 4.0533 & -0.0090 & 0.0720 & 0.0744          & 0.0745        & 0.0845 & 93.1\% &  & 3.4661 & 0.0038  & 0.0899 & 0.0882          & 0.0882        & 0.0991 & 90.4\%  \\
50   & 1    & Low         & Yes   & 4.0533 & -0.0509 & 0.0719 & 0.0823          &               & 0.0848 & 93.4\% &  & 3.4661 & -0.0386 & 0.0774 & 0.0820          &               & 0.0731 & 90.8\%  \\
50   & 3    & Low         & Yes   & 4.0533 & -0.0009 & 0.0456 & 0.0470          & 0.0470        & 0.0340 & 94.1\% &  & 3.4661 & 0.0039  & 0.0569 & 0.0562          & 0.0563        & 0.0393 & 92.5\%  \\
100  & 1    & Low         & Yes   & 4.0533 & -0.0242 & 0.0510 & 0.0566          &               & 0.0428 & 94.3\% &  & 3.4661 & -0.0198 & 0.0556 & 0.0572          &               & 0.0373 & 91.7\%  \\
100  & 3    & Low         & Yes   & 4.0533 & -0.0004 & 0.0324 & 0.0330          & 0.0330        & 0.0172 & 94.5\% &  & 3.4661 & 0.0043  & 0.0400 & 0.0399          & 0.0399        & 0.0193 & 93.5\%  \\
500  & 1    & Low         & Yes   & 4.0533 & -0.0021 & 0.0232 & 0.0242          &               & 0.0089 & 95.0\% &  & 3.4661 & -0.0020 & 0.0249 & 0.0254          &               & 0.0075 & 94.1\%  \\
500  & 3    & Low         & Yes   & 4.0533 & 0.0021  & 0.0144 & 0.0147          & 0.0147        & 0.0034 & 95.2\% &  & 3.4661 & 0.0020  & 0.0177 & 0.0178          & 0.0179        & 0.0038 & 94.7\%  \\
\hline
\end{tabular}
\begin{tablenotes}\footnotesize
\item [a] Standard deviation of 10,000 estimated OTDIs which should be close to the true standard error of the estimator
\item [b] Mean estimated standard error of estimators from GEE with independent correlation matrix
\item [c] Mean estimated standard error of estimators from GEE with true correlation matrix (This value is left blank when there is no replicates because it is the same as $SE_{ind}$)
\end{tablenotes}
  \end{threeparttable}
\end{sidewaystable}

\begin{sidewaystable}
\footnotesize
\centering
\begin{threeparttable}
\caption{Simulation Results of Relative Area under Overall Coverage Probability Curve}
\label{tab::simuout_r}
  \begin{tabular}{rrrrrrrrrrrrrrrrrrr}
\hline
\multicolumn{1}{c}{\multirow{2}{*}{nsub}} & \multicolumn{1}{c}{\multirow{2}{*}{nrep}} & \multicolumn{1}{c}{\multirow{2}{*}{Corr}} & \multicolumn{1}{c}{\multirow{2}{*}{Shift}} & \multicolumn{7}{c}{Normal} & \multicolumn{1}{c}{} & \multicolumn{7}{c}{Lognormal} \\ \cline{5-11} \cline{13-19} 
&&&&\multicolumn{1}{c}{True}&&&&&&&&\multicolumn{1}{c}{True}\\
\multicolumn{1}{c}{} & \multicolumn{1}{c}{} & \multicolumn{1}{c}{} & \multicolumn{1}{c}{} & \multicolumn{1}{c}{RAUOCPC} & \multicolumn{1}{c}{Bias} & \multicolumn{1}{c}{$SD_{t}$\tnote{a}} & \multicolumn{1}{c}{$SE_{ind}$\tnote{b}} & \multicolumn{1}{c}{$SE_{t}$\tnote{b}} & \multicolumn{1}{c}{MSE} & \multicolumn{1}{c}{CR} & \multicolumn{1}{c}{} & \multicolumn{1}{c}{RAUOCPC} & \multicolumn{1}{c}{Bias} & \multicolumn{1}{c}{$SD_{t}$\tnote{a}} & \multicolumn{1}{c}{$SE_{ind}$\tnote{b}} & \multicolumn{1}{c}{$SE_{t}$\tnote{c}} & \multicolumn{1}{c}{MSE} & \multicolumn{1}{c}{CR} \\ \hline
20 & 1 & High & No & 0.6084 & -0.0004 & 0.1923 & 0.1910 &    & 0.0021 & 95.1\% &  & 0.7720 & 0.0002 & 0.2996 & 0.2835 &    & 0.0027 & 96.2\% \\
20 & 3 & High & No & 0.6084 & -0.0008 & 0.1397 & 0.1368 & 0.1477 & 0.0011 & 93.5\% &  & 0.7720 & -0.0058 & 0.2460 & 0.2543 & 0.2549 & 0.0020 & 95.1\% \\
50 & 1 & High & No & 0.6084 & -0.0002 & 0.1219 & 0.1215 &    & 0.0008 & 95.4\% &  & 0.7720 & -0.0002 & 0.1859 & 0.1839 &    & 0.0011 & 96.0\% \\
50 & 3 & High & No & 0.6084 & -0.0002 & 0.0902 & 0.0886 & 0.0981 & 0.0005 & 93.7\% &  & 0.7720 & -0.0001 & 0.1679 & 0.1620 & 0.1627 & 0.0009 & 92.5\% \\
100 & 1 & High & No & 0.6084 & -0.0001 & 0.0855 & 0.0860 &    & 0.0004 & 95.1\% &  & 0.7720 & 0.0000 & 0.1323 & 0.1309 &    & 0.0005 & 95.6\% \\
100 & 3 & High & No & 0.6084 & -0.0001 & 0.0633 & 0.0631 & 0.0710 & 0.0002 & 94.1\% &  & 0.7720 & -0.0004 & 0.1179 & 0.1159 & 0.1166 & 0.0004 & 93.9\% \\
500 & 1 & High & No & 0.6084 & 0.0000 & 0.0382 & 0.0385 &    & 0.0001 & 95.6\% &  & 0.7720 & 0.0000 & 0.0588 & 0.0588 &    & 0.0001 & 95.4\% \\
500 & 3 & High & No & 0.6084 & 0.0000 & 0.0281 & 0.0284 & 0.0331 & 0.0000 & 95.1\% &  & 0.7720 & 0.0000 & 0.0522 & 0.0523 & 0.0526 & 0.0001 & 94.5\% \\ \hline
20 & 1 & Low & No & 0.4911 & -0.0006 & 0.2288 & 0.2280 &    & 0.0032 & 95.6\% &  & 0.6784 & -0.0001 & 0.2790 & 0.2718 &    & 0.0036 & 96.0\% \\
20 & 3 & Low & No & 0.4911 & -0.0005 & 0.1448 & 0.1403 & 0.1409 & 0.0013 & 93.5\% &  & 0.6784 & -0.0003 & 0.2172 & 0.2083 & 0.2090 & 0.0022 & 92.2\% \\
50 & 1 & Low & No & 0.4911 & -0.0004 & 0.1424 & 0.1436 &    & 0.0013 & 95.3\% &  & 0.6784 & -0.0003 & 0.1721 & 0.1733 &    & 0.0014 & 95.8\% \\
50 & 3 & Low & No & 0.4911 & -0.0002 & 0.0907 & 0.0903 & 0.0908 & 0.0005 & 94.1\% &  & 0.6784 & 0.0000 & 0.1390 & 0.1346 & 0.1350 & 0.0009 & 93.3\% \\
100 & 1 & Low & No & 0.4911 & -0.0002 & 0.1009 & 0.1014 &    & 0.0006 & 95.2\% &  & 0.6784 & 0.0001 & 0.1238 & 0.1227 &    & 0.0007 & 95.5\% \\
100 & 3 & Low & No & 0.4911 & -0.0001 & 0.0645 & 0.0643 & 0.0646 & 0.0003 & 94.2\% &  & 0.6784 & -0.0003 & 0.0972 & 0.0959 & 0.0962 & 0.0004 & 94.4\% \\
500 & 1 & Low & No & 0.4911 & 0.0000 & 0.0450 & 0.0453 &    & 0.0001 & 95.4\% &  & 0.6784 & 0.0000 & 0.0547 & 0.0549 &    & 0.0001 & 95.5\% \\
500 & 3 & Low & No & 0.4911 & 0.0000 & 0.0288 & 0.0289 & 0.0290 & 0.0001 & 94.9\% &  & 0.6784 & 0.0000 & 0.0427 & 0.0431 & 0.0433 & 0.0001 & 94.7\% \\ \hline
20 & 1 & High & Yes & 0.3583 & -0.0005 & 0.2347 & 0.2348 &    & 0.0028 & 95.0\% &  & 0.3943 & -0.0002 & 0.1724 & 0.1716 &    & 0.0017 & 96.3\% \\
20 & 3 & High & Yes & 0.3583 & -0.0004 & 0.1936 & 0.1886 & 0.1912 & 0.0019 & 94.5\% &  & 0.3943 & -0.0001 & 0.1504 & 0.1444 & 0.1457 & 0.0013 & 92.7\% \\
50 & 1 & High & Yes & 0.3583 & -0.0001 & 0.1489 & 0.1479 &    & 0.0012 & 95.0\% &  & 0.3943 & 0.0000 & 0.1083 & 0.1087 &    & 0.0007 & 96.0\% \\
50 & 3 & High & Yes & 0.3583 & -0.0002 & 0.1230 & 0.1210 & 0.1227 & 0.0008 & 94.5\% &  & 0.3943 & -0.0004 & 0.0947 & 0.0930 & 0.0940 & 0.0005 & 94.0\% \\
100 & 1 & High & Yes & 0.3583 & -0.0001 & 0.1038 & 0.1044 &    & 0.0006 & 95.1\% &  & 0.3943 & 0.0002 & 0.0760 & 0.0769 &    & 0.0003 & 95.9\% \\
100 & 3 & High & Yes & 0.3583 & -0.0001 & 0.0862 & 0.0859 & 0.0872 & 0.0004 & 94.5\% &  & 0.3943 & -0.0003 & 0.0668 & 0.0662 & 0.0669 & 0.0003 & 94.2\% \\
500 & 1 & High & Yes & 0.3583 & 0.0000 & 0.0462 & 0.0466 &    & 0.0001 & 95.4\% &  & 0.3943 & 0.0000 & 0.0343 & 0.0344 &    & 0.0001 & 95.2\% \\
500 & 3 & High & Yes & 0.3583 & 0.0000 & 0.0383 & 0.0385 & 0.0391 & 0.0001 & 94.8\% &  & 0.3943 & 0.0000 & 0.0298 & 0.0297 & 0.0300 & 0.0001 & 94.7\% \\ \hline
20 & 1 & Low & Yes & 0.3100 & 0.0003 & 0.2772 & 0.2774 &    & 0.0033 & 95.3\% &  & 0.3559 & 0.0000 & 0.2174 & 0.2166 &    & 0.0024 & 96.0\% \\
20 & 3 & Low & Yes & 0.3100 & -0.0005 & 0.1967 & 0.1917 & 0.1921 & 0.0017 & 94.6\% &  & 0.3559 & 0.0001 & 0.1731 & 0.1661 & 0.1667 & 0.0015 & 93.6\% \\
50 & 1 & Low & Yes & 0.3100 & -0.0001 & 0.1746 & 0.1740 &    & 0.0014 & 95.0\% &  & 0.3559 & 0.0001 & 0.1351 & 0.1363 &    & 0.0009 & 95.5\% \\
50 & 3 & Low & Yes & 0.3100 & -0.0001 & 0.1249 & 0.1230 & 0.1232 & 0.0007 & 94.5\% &  & 0.3559 & -0.0005 & 0.1090 & 0.1066 & 0.1070 & 0.0006 & 94.3\% \\
100 & 1 & Low & Yes & 0.3100 & 0.0000 & 0.1222 & 0.1226 &    & 0.0007 & 95.0\% &  & 0.3559 & 0.0003 & 0.0956 & 0.0963 &    & 0.0005 & 95.5\% \\
100 & 3 & Low & Yes & 0.3100 & -0.0001 & 0.0876 & 0.0873 & 0.0875 & 0.0003 & 94.4\% &  & 0.3559 & -0.0003 & 0.0765 & 0.0759 & 0.0761 & 0.0003 & 94.5\% \\
500 & 1 & Low & Yes & 0.3100 & 0.0000 & 0.0552 & 0.0547 &    & 0.0001 & 95.0\% &  & 0.3559 & 0.0000 & 0.0429 & 0.0430 &    & 0.0001 & 95.2\% \\
500 & 3 & Low & Yes & 0.3100 & 0.0000 & 0.0389 & 0.0391 & 0.0392 & 0.0001 & 94.9\% &  & 0.3559 & 0.0000 & 0.0339 & 0.0340 & 0.0341 & 0.0001 & 94.8\% \\ \hline
\end{tabular}
\begin{tablenotes}\footnotesize
\item [a] Standard deviation of 10,000 estimated RAUOCPC which is expected to be the true standard error of the estimator for a very large number of simulations 
\item [b] Mean estimated standard error of estimators from GEE with independent correlation matrix
\item [c] Mean estimated standard error of estimators from GEE with true correlation matrix (This value is left blank when there is no replicates due to no difference from $SE_{ind}$)
\end{tablenotes}
    \end{threeparttable}
\end{sidewaystable}

\section{BP Example}\label{sec::BP}
The proposed indices and inference approach are illustrated with the systolic blood pressure data in the Bland and Altman's paper \cite{bland1986}. In this data example, the blood pressures of 85 patients were measured by three raters (two human observers J and R and one device S). Each raters measured every patients three times successively that can be treated as replicates.We assess the overall agreement among these three raters along with the intra-rater agreement within each raters as well as the pairwise inter agreement by OCP, OTDI and ORAUCPC with estimation and inference conducted by the proposed unified GEE approach. 

The descriptive statistics of BP data is listed in table (\ref{tab::bp_ss}). We summarize the data by mean and stander deviation within each raters and assess the normality assumption of replicates from same rater and pairwise difference between any two raters by Doornik-Hansen's test where a p-value less than 0.05 indicates a significant departure form a multivariate normal distribution. As shown in the Table \ref{tab::bp_ss} , the human rater S tends to have higher BP measurements with a average measurement of 143.04 mmHg than the other two raters whose numbers are around 127 mmHg. Moreover, based on $SD_{intra}$, the rater S has lager within-rater variability than the raters J and R which implies that the heterogeneity among the raters exists. Furthermore, the p-value of Doornik-Hansen's test for the measurement from each raters and the difference between raters are all less than 0.05 indicating that the normality assumption required for the estimation and inference approaches of unscaled indexes proposed by Lin\cite{lin2007unified} and Jang et al.\cite{jang2018overall} do not hold and their approaches are unsuitable for the BP dataset.

To assess the overall agreement among three raters, the new OCP, OTDI and RAUOCPC are used in analyzing the BP dataset. The satisfactory agreement is set based on the British hypertension society protocol (BHSP) for the evaluation of blood pressure measuring device \cite{o1993british} shown in Table \ref{tab::bp_ss}. For OCP, we set the pre-determined clinically meaningful acceptable difference to be $\delta_{0}=15$mmHg and based on the criteria for grade C device the corresponding satisfactory OCP should be $0.85$ or higher. For OTDI, pre-determined acceptable probability is set to be $\pi_{0}=0.85$ and the satisfactory OTDI for grade C device is 15mmHg. For RAUOCPC, let $\delta_{max}=20$mmHg and the satisfactory RAUOCPC is 0.59 which is computed based on overall coverage probability curve that connect points formed by the absolute differences of 0, 5, 10, 15, and 20 with the corresponding coverage probabilities for BP device of grade C with $\delta_{max}=20$. 

\begin{figure}[ht!]
\centering
\includegraphics[scale=0.6]{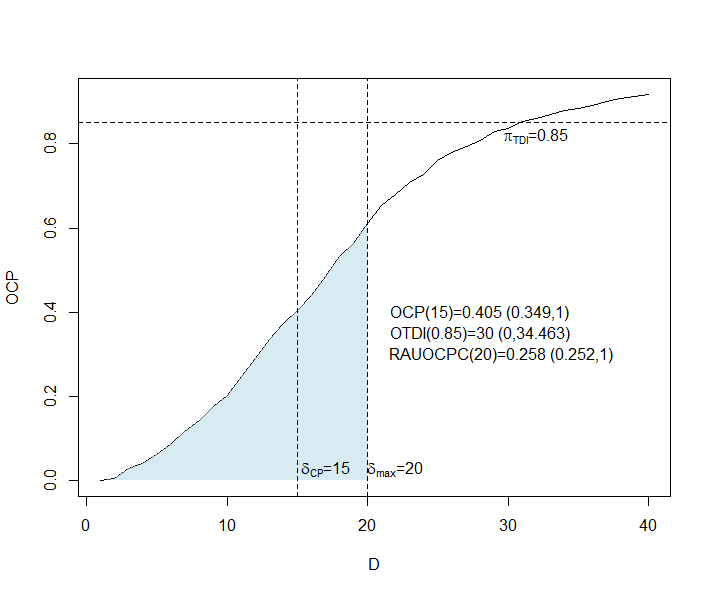}
\caption{Overall coverage probability curve for BP data set}
\label{fig::bpcp}
\end{figure}

The estimated coverage probability curve is shown in Figure \ref{fig::bpcp}. The estimated OCP is 0.41 with 95\% one-sided CI of (0.35,1) for three raters. Since the CI contains $\pi_0=0.8$, we cannot reject the null hypothesis and thus there is no sufficient evidence to claim that three raters can be used interchangeably. We can come to the same conclusion with OTDI and RAUOCPC. The estimated OTDI is 30 with 95\% one-sided CI of (0, 34.5) which contains $\delta_0=15$ and estimated RAUOCPC is 0.258 with 95\% one-sided CI of (0.25,1) which contains $\tau_0=0.59$. Therefore, based on the proposed overall agreement, three raters may not be used interchangeably in the sense that we are not confident that the measurements taken by three raters on the same patients are clinically similar.

To understand the source of disagreement and provide actionable results that guide readers to improve quality, we look into the pairwise inter-rater and intra-rater agreement between and within three raters, respectively. The results listed in Table \ref{tab::bp_ss} show that both the intra-rater agreement of human raters J and R and the inter-rater agreement between them are satisfactory. This implies that two human raters can be used interchangeably and the measurements from different nurses or different replicates from the same nurse are not likely to be clinically different. However, the agreement between human nurses and the deceive S is less satisfactory where the inter-rater OCPs (one sided 95\% CI) are 0.51(0.45,1) and 0.51(0.45,1). Moreover, the repeatability of device S itself is also moderate with estimated intra-rater OCP of 0.84(0.78,1) and OTDI of 15(0,17.32). These results indicate that not only the device S is not in satisfactory agreement with the other raters but also its own replicates tend to have larger variability.

\begin{table}[ht!]
\centering
\caption{Descriptive statistics for BP data}
\label{tab::bp_ss}
\begin{tabular}{cccc}
\hline
Rater & Mean  & $SD_{intra}$ & P-Value for Normality Test \\ \hline
J   & 127.4 & 5.3    & 0.004           \\
R   & 127.3 & 5.4    & 0.008           \\
S   & 143.0 & 7.0    & \textless{}0.001      \\ \hline
\end{tabular}
\end{table}

\begin{table}[ht!]
\centering
\captionof{table}{Intra, Inter and Overall Agreement for Three Raters}\label{table::bp}
\begin{tabular}{llllllllll}
\hline
\multicolumn{2}{l}{\multirow{2}{*}{}} & \multicolumn{2}{c}{CP} &  & \multicolumn{2}{c}{TDI}      &  & \multicolumn{2}{c}{RAUCPC} \\ \cline{3-4} \cline{6-7} \cline{9-10} 
\multicolumn{2}{l}{}                  & Estimation & 95\% CI   &  & Estimation & 95\% CI         &  & Estimation   & 95\% CI     \\ \hline
Overall                     &         & 0.41      & (0.35, 1) &  & 30         & (0, 34.46) &  & 0.26        & (0.25, 1)   \\
\multirow{3}{*}{Inter}     & J\&R    & 0.94      & (0.91, 1) &  & 10         & (0, 10.89)  &  & 0.76        & (0.74, 1)   \\
                            & J\&S    & 0.51      & (0.45, 1) &  & 28         & (0, 32.47) &  & 0.34        & (0.33, 1)   \\
                            & R\&S    & 0.51      & (0.45,  1) &  & 28        & (0, 32.31) &  & 0.35        & (0.34, 1)   \\
\multirow{3}{*}{Intra}     & J       & 0.91      & (0.87, 1) &  & 12         & (0, 13.48)  &  & 0.67        & (0.65, 1)   \\
                            & R       & 0.92      & (0.88, 1) &  & 13         & (0, 14.21) &  & 0.66        & (0.65, 1)   \\
                            & S       & 0.84      & (0.78, 1) &  & 15         & (0,  17.32) &  & 0.60        & (0.59, 1)   \\ \hline
\end{tabular}
\end{table}

\section{Discussion}
We have proposed a set of new indices (OCP, OTDI and RAUOCPC) for assessing overall agreement among among multiple raters. As an extension from the pairwise version of unscaled indices, the proposed indices are defined based on a new distance metric which measures the maximum pairwise difference among the raters. This metric allows the overall indices to preserve the intuitive interpretation from the pairwise version and directly employs the clinically information about satisfactory criteria. For example, we can extend clinically meaningful difference from the grading system of blood pressure device as the pre-determine boundary for OCP, since they both quantify the acceptable difference between two BP measurements. The OCP can be interpreted as the probability there is no clinically meaningful difference among measurements from all raters on the same subject. 

The new proposed inference approach does not require distributional and homogeneity assumptions and therefore can be applied to various kinds of continuous measurements. As we discuss in Section \ref{sec::BP}, the BP data set \cite{bland1986} is neither homogeneous nor normally distributed which are the assumptions the previously proposed inference approach. Moreover, the unified GEE approach could accommodate data with replicates and can be easily modified to carry out estimation and inference on pairwise, inter-rater and intra-rater agreements as we did in the BP example. The design with replicates is preferable since it can provide information on the repeatability of the raters. When the agreement is not satisfactory, intra-rater variability is a crucial source of disagreement and such information could provide guideline for future improvement of the testing raters. In addition to provide additional information, adding replication also could improve the performance of the estimator in terms of bias and CR as shown in our simulation studies (Table \ref{tab::simuout_cp}, \ref{tab::simuout_tdi} and \ref{tab::simuout_r}). In practice, it tends to be easier and less costly by adding replicates than enrolling more subjects. 

All proposed estimation and inference approaches can be easily applied by standard software and we also provide the R package for implementation. Based on the simulation results, the proposed approaches have limitation when the sample size is small and no replicates are available. For such scenario, parametric approaches can be an alternative after carefully verifying the assumptions. Moreover, it is of future interest to design agreement study based on the new proposed indices especially for design with replicates. As we discuss before, adding replicates can provide information on intra-rater agreement and improve the performance of estimators.

%\bibliography{references} %%% Remove comment to use the external .bib file (using bibtex).


\bibliographystyle{unsrt}  
\bibliography{references}

\begin{thebibliography}{10}

\bibitem{barnhart2007overview}
Huiman~X Barnhart, Michael~J Haber, and Lawrence~I Lin.
\newblock An overview on assessing agreement with continuous measurements.
\newblock {\em Journal of biopharmaceutical statistics}, 17(4):529--569, 2007.

\bibitem{barnhart2016choice}
Huiman~X Barnhart, Eric Yow, Anna~Lisa Crowley, Melissa~A Daubert, Dawn
  Rabineau, Robert Bigelow, Michael Pencina, and Pamela~S Douglas.
\newblock Choice of agreement indices for assessing and improving measurement
  reproducibility in a core laboratory setting.
\newblock {\em Statistical methods in medical research}, 25(6):2939--2958,
  2016.

\bibitem{o1993british}
Eoin O’Brien, James Petrie, WA~Littler, Michael de~Swiet, Paul~L Padfield,
  Douglas Altman, Martin Bland, Andrew Coats, Neil Atkins, et~al.
\newblock The british hypertension society protocol for the evaluation of blood
  pressure measuring devices.
\newblock {\em J hypertens}, 11(Suppl 2):S43--S62, 1993.

\bibitem{raucpc}
Huiman~X. Barnhart.
\newblock Assessing agreement with relative area under the coverage probability
  curve.
\newblock {\em Statistics in Medicine}, 35(18):3153--3165, 2016.

\bibitem{bland1986}
J~Martin Bland, Douglas~G Altman, et~al.
\newblock Statistical methods for assessing agreement between two methods of
  clinical measurement.
\newblock {\em lancet}, 1(8476):307--310, 1986.

\bibitem{lin2007unified}
Lawrence Lin, AS~Hedayat, and Wenting Wu.
\newblock A unified approach for assessing agreement for continuous and
  categorical data.
\newblock {\em Journal of Biopharmaceutical Statistics}, 17(4):629--652, 2007.

\bibitem{jang2018overall}
Jeong~Hoon Jang, Amita~K Manatunga, Andrew~T Taylor, and Qi~Long.
\newblock Overall indices for assessing agreement among multiple raters.
\newblock {\em Statistics in medicine}, 37(28):4200--4215, 2018.

\bibitem{liang1986longitudinal}
Kung-Yee Liang and Scott~L Zeger.
\newblock Longitudinal data analysis using generalized linear models.
\newblock {\em Biometrika}, 73(1):13--22, 1986.

\bibitem{duong2019package}
Tarn Duong, Maintainer~Tarn Duong, and MASS Suggests.
\newblock Package ‘ks’.
\newblock 2019.

\bibitem{wang2005effects}
You-Gan Wang and Xu~Lin.
\newblock Effects of variance-function misspecification in analysis of
  longitudinal data.
\newblock {\em Biometrics}, 61(2):413--421, 2005.

\bibitem{wang2003working}
You-Gan Wang and Vincent Carey.
\newblock Working correlation structure misspecification, estimation and
  covariate design: implications for generalised estimating equations
  performance.
\newblock {\em Biometrika}, 90(1):29--41, 2003.

\bibitem{jung1996quasi}
Sin-Ho Jung.
\newblock Quasi-likelihood for median regression models.
\newblock {\em Journal of the American Statistical Association},
  91(433):251--257, 1996.

\bibitem{halliwell2015lognormal}
Leigh~J Halliwell.
\newblock The lognormal random multivariate.
\newblock In {\em Casualty Actuarial Society E-Forum, Spring}, volume~5, 2015.

\end{thebibliography}

\newpage
\begin{appendices}

\section{Proof of Lemma \ref{lemma::raucpc}}\label{app::raucpc}
Let $C=\max(0,\delta_{max}-D)\in[0,\delta_{max}]$,  and the cumulative distribution functions of $C$ and $D$ be $F_{C}(c)$ and $F_{D}(d)$ and the density functions be $f_{C}(c)$ and $f_{D}(d)$ respectively. Then,
 \begin{align}
     F_C(c)&=Pr(C\leq c)\\
     &=Pr(\max(0,\delta_{max}-D) \leq c)\\
     &=
     \begin{cases}
        Pr(D\geq c-\delta_{max})   & \quad \text{if } 0 < c\leq\delta_{max}\\
        Pr(D\geq\delta_{max})   & \quad \text{if } c = 0
  \end{cases}\\
  &=\begin{cases}
        1-F_D(\delta_{max}-c)^-   & \quad \text{if } 0 < c \leq\delta_{max}\\
        1-F_D(\delta_{max})  & \quad \text{if } c =0
  \end{cases}
 \end{align}
 Thus,
\begin{align}
     f_C(c)&=\begin{cases}
        f_D(\delta_{max}-c)   & \quad \text{if } 0 < c \leq\delta_{max}\\
        P(c=\delta_{max})  & \quad \text{if } c = 0
  \end{cases}\\
  &=\begin{cases}
          f_D(\delta_{max}-c)    & \quad \text{if }  0 < c \leq\delta_{max}\\
        1-F_D(\delta_{max})  & \quad \text{if } c = 0
  \end{cases}
 \end{align}
Therefore,
\begin{align}
E[C]=&\int_{0}^{\delta_{max}}cf_C(c)\mathrm{d}c\\
=&\int_{0^+}^{\delta_{max}}c  f_D(\delta_{max}-c) \mathrm{d}c+0\times [1-F_D(\delta_{max})]\\
=&\int_{0}^{\delta_{max}^-} (\delta_{max}-d)  f_D(d) \mathrm{d}d \label{equ::raucpc00}
\end{align}
Since $f_D(d)$ is continuous at point $\delta_{max}$, then
\begin{align}
\\
(\ref{equ::raucpc00})=&\int_{0}^{\delta_{max}} (\delta_{max}-d)  f_D(d) \mathrm{d}d\\
=& (\delta_{max}-d) F_D(d)|_{0}^{\delta_{max}}+\int_{0}^{\delta_{max}}F_D(d)\mathrm{d}d\\
=&\int_{0}^{\delta_{max}}df_D(d)\mathrm{d}d+\delta_{max}[1-F_X(\delta_{max})]\\
=&\int_{0}^{\delta_{max}}F_D(d)\mathrm{d}d\label{equ::raucpc0}
\end{align}

This implies that $\int_{0}^{\delta_{max}}F_D(x)dx=E[C]$ and therefor  RAUCPC can be expressed in terms of $E(C)$.
\begin{align}
   RAUCPC(\delta_{max})&=\frac{\int_{0}^{\delta_{max}}CP(d)\mathrm{d}d}{\delta_{max}}\\
   &= \frac{\int_{0}^{\delta_{max}}F_D(d)\mathrm{d}d}{\delta_{max}}\label{equ::raucpc1}\\
   &=\frac{{E}[C]}{\delta_{max}}
\end{align}

\section{Proof of asymptotic distribution in estimating OTDI} \label{app::OTDI}

In equation (19) for estimating OTDI, we propose to use $\Gamma_i=f_D(g(\theta))g'(\theta)\mathbf{1}_{M\times 1}$. Let $H_i(\theta)=\Gamma_i^T V_i$ where $V_i=A_i^{1/2}R_aA_i^{1/2}$ with working correlation matrix $R_a$. Then the left hand side of (\ref{equ::gee0}) is $U_N(\theta)=\sum_{i=1}^N H_i(\theta)^TS_i(\theta)$  where $S_i(\theta)=\left(\pi_0-I(D_{i1}-g(\theta)<0),\cdots, \pi_0-I(D_{iM}-g(\theta)<0)\right)$. Let $\bar{U}_N(\theta)=\sum_{i=1}^N H_i(\theta)\bar{S}_i(\theta)$ where $\bar{S}_i(\theta)=\left(\pi_0-\Pr(D_{i1}-g(\theta)<0),\cdots.\pi_0-\Pr(D_{iM}-g(\theta)<0)\right)$. Then 
\begin{align}
    N^{-1}\left\{U_N(\theta)- \bar{U}_N(\theta)\right\}=&N^{-1}\sum_{i=1}^N H_i(\theta)(S_i-\bar{S}_i)\\
    =&N^{-1}\sum_{i=1}^N H_i(\theta) \begin{pmatrix}
    \Pr(D_{i1}-g(\theta)<0)-I(D_{i1}-g(\theta)<0)\\
    \vdots\\
    \Pr(D_{iM}-g(\theta)<0)-I(D_{iM}-g(\theta)<0)
    \end{pmatrix}
\end{align}
Under mild regulations, by uniform strong law of large
numbers\cite{jung1996quasi}, we have
\begin{align}
    \sup_{\theta \in \Theta}\left| N^{-1}\left\{U_N(\theta)- \bar{U}_N(\theta)\right\}\right| \rightarrow o(N^{-1/2}) \quad a.s \label{equ::suptheta}
\end{align}Then we can write $U_N(\theta)$ as
\begin{align}
N^{-1/2}U_N(\theta)&=N^{-1/2}\bar{U}_N(\theta)+o(1)\\
&=N^{-1/2}\left\{U_N(\theta_0)-U_N(\theta_0)\right\}+N^{-1/2}\left\{\bar{U}_N(\theta_0)+\frac{\partial}{\partial \theta} \bar{U}_N(\theta_0) (\theta-\theta_0)\right\}+ o(1)\\
&=N^{-1/2}\left\{U_N(\theta_0)+\frac{\partial}{\partial \theta} \bar{U}_N(\theta_0) (\theta-\theta_0)\right\}+N^{-1/2}\left\{\bar{U}_N(\theta_0)-U_N(\theta_0)\right\}+ o(1)
\end{align}
Suppose $\hat{\theta}$ is the solution of $U_N(\theta)=0$ such that $U_N(\hat{\theta})=0$. With $N^{-1/2}\left\{\bar{U}_N(\theta_0)-U_N(\theta_0)\right\}\rightarrow o(1)$ from (\ref{equ::suptheta}), then
\begin{align}
N^{1/2}(\hat{\theta}-\theta_0)=-N^{-1/2}A_N^{-1}U_N(\theta_0)+o(1)
\end{align}
where,
\begin{align}
    A_N=N^{-1}\frac{\partial}{\partial \theta} \bar{U}_N(\theta_0) \rightarrow A=\lim_{N\rightarrow \infty} N^{-1}\sum_{i=1}^NH_i(\theta_0)\begin{pmatrix}
    f_1(g(\theta_0))g'(\theta_0)\\
    \vdots\\
     f_M(g(\theta_0))g'(\theta_0)
    \end{pmatrix}
\end{align}
where $f_m(\cdot)$ is the density function of $D_{im}$and since all $D_{im}$ follows the same marginal distribution we denote their common density function as $f_D(\cdot)$.  Thus, $A=\lim_{N\rightarrow \infty} N^{-1}\sum_{i=1}^N\Gamma_i^TV_i\Gamma_i$. 
By central limit theorem,
\begin{align}
N^{-1/2}U_N(\theta_0)\rightarrow \mathcal{N}(0,\Sigma_0)
\end{align}
where $\Sigma_0=\sum_{i=1}^N  \Gamma_i^T V_i^{-1}cov(S_i)V_i^{-1}\Gamma_i$.Then,

\begin{align}
N^{1/2}(\hat{\theta}-\theta_0)\rightarrow \mathcal{N}(0,V_{\theta}^{sw})
\end{align}
where $V_{\theta}^{sw}=\lim_{N\rightarrow \infty}N\Sigma_1^{-1}\Sigma_0\Sigma_1^{-1}$, $\Sigma_1=\sum_{i=1}^N \Gamma_i^T V_i^{-1}\Gamma_i$ and $\Sigma_0=\sum_{i=1}^N  \Gamma_i^T V_i^{-1}cov(S_i)V_i^{-1}\Gamma_i$.
\section{Proof of Lemma \ref{lemma::euqalrow}}\label{app::eqeualrow} 
We first prove the Lemma for the case with three raters ($J=3$) and two replications ($K=2$) and then extend the proof to the general case.  Denote $D_{im}=max(|x_{i1k_1}-x_{i2k_2}|,|x_{i1k_1}-x_{i3k_3}|,|x_{i2k_2}-x_{i3k_3}|))$ and $D_{im'}=max(|x_{i1k_1'}-x_{i2k_2'}|,|x_{i1k_1'}-x_{i3k_3'}|,|x_{i2k_2'}-x_{i3k_3'}|),\theta)$ as two maximum distances among the three raters. Define $\mathcal{O}_{mm'}$ as a collection of index of the raters whose replicates are common in the two terms i.e. $\mathcal{O}_{mm'}=\{j;k_j=k_j'\}$. 
 Since we assume that replicates from same rater on same subject are iid,  then the  correlation between $s(D_{im},\theta)$ and $s(D_{im'},\theta)$ is different only up to the common reads of $x_{ijk}$'s used in  $s(D_{im},\theta)$ and $s(D_{im'},\theta)$ indexed by $\mathcal{O}_{mm'}$. Specifically
\begin{itemize}
\item If $k_1\neq k_1'$, $k_2 \neq k_2'$ and $k_3 \neq k_3'$, then $\mathcal{O}_{mm'}=\emptyset$ where $s(D_{im},\theta)$ and $s(D_{im'},\theta)$ contain zero common reads from three raters between them and we use index 0 to denote such correlation $corr(s(D_{im},\theta),s(D_{im'},\theta))=\rho_0$
\item If $k_1= k_1'$, $k_2 \neq k_2'$ and $k_3 \neq k_3'$, then $\mathcal{O}_{mm'}=\{1\}$ where $s(D_{im},\theta)$ and $s(D_{im'},\theta)$ contain exactly one common read from the first rater and we use index $1=2^{1-1}$ to denote such correlation $corr(s(D_{im},\theta),s(D_{im'},\theta))=\rho_1$
\item If $k_1\neq k_1'$, $k_2 = k_2'$ and $k_3 \neq k_3'$, then $\mathcal{O}_{mm'}=\{2\}$, then $s(D_{im},\theta)$ and $s(D_{im'},\theta)$ contain exactly one common read from the second rater and we used index $2=2^{2-1}$ to denote $corr(s(D_{im},\theta),s(D_{im'},\theta))=\rho_2$
\item If $k_1= k_1'$, $k_2 = k_2'$ and $k_3 \neq k_3'$, then $\mathcal{O}_{mm'}=\{1,2\}$ where $s(D_{im},\theta)$ and $s(D_{im'},\theta)$ contain exactly two common reads from the first rater and second rater and we used index $3=2^{1-1}+2^{2-1}$ to denote $corr(s(D_{im},\theta),s(D_{im'},\theta))=\rho_3$
\item If $k_1\neq k_1'$, $k_2 \neq k_2'$ and $k_3 = k_3'$, then $\mathcal{O}_{mm'}=\{3\}$ where $s(D_{im},\theta)$ and $s(D_{im'},\theta)$ contain exactly one common read from the third rater and we used index $4=2^{3-1}$ to denote $corr(s(D_{im},\theta),s(D_{im'},\theta))=\rho_4$
\item If $k_1= k_1'$, $k_2 \neq k_2'$ and $k_3 =k_3'$, then $\mathcal{O}_{mm'}=\{1,3\}$ where $s(D_{im},\theta)$ and $s(D_{im'},\theta)$ contain exactly two common reads from the first rater the third rater and we use index $5=2^{1-1}+2^{3-1}$ to denote $corr(s(D_{im},\theta),s(D_{im'},\theta))=\rho_5$
\item If $k_1\neq k_1'$, $k_2 = k_2'$ and $k_3 = k_3'$, then $\mathcal{O}_{mm'}=\{2,3\}$ where $s(D_{im},\theta)$ and $s(D_{im'},\theta)$ contain exactly two common reads from the second rater the third rater and we use index $6=2^{2-1}+2^{3-1}$ to denote $corr(s(D_{im},\theta),s(D_{im'},\theta))=\rho_6$
\item If $k_1=k_1'$, $k_2=k_2'$ and $k_3=k_3'$, then $\mathcal{O}_{mm'}=\{1,2,3\}$ where $s(D_{im},\theta)=s(D_{im'},\theta)$ contain exactly three common reads for all three raters and we use index $7=2^{1-1}+2^{2-1}+2^{3-1}$ to denote $corr(s(D_{im},\theta),s(D_{im'},\theta))=\rho_7=1$
\end{itemize}

With these notations for the $\begin{pmatrix}3\\0\end{pmatrix}+\begin{pmatrix}3\\1\end{pmatrix}+\begin{pmatrix}3\\2\end{pmatrix}+\begin{pmatrix}3\\3\end{pmatrix}=2^3$ distinct correlations where correlation is denoted as $\rho_l, l=\sum_{j\in \mathcal{O}_{mm'}} 2^{j-1}$, the true correlation matrix can be written as, 
\begin{align*}
\small
 \mathbf{R}_0=\bordermatrix{~ & s_{111}& s_{112} & s_{121} &  s_{122}& s_{211} & s_{212} & s_{221}& s_{222}  \cr
		s_{111} & 1 & \rho_3 & \rho_5 & \rho_1 & \rho_6 & \rho_2 & \rho_4 &\rho_0   \cr
		s_{112} & \rho_3 & 1 &\rho_1 & \rho_5  & \rho_2 & \rho_6  & \rho_0  & \rho_4  \cr
        s_{121} & \rho_5 & \rho_1 &1 & \rho_3  & \rho_4 & \rho_0  & \rho_6   & \rho_2 \cr
        s_{122} & \rho_1 & \rho_5 & \rho_3 & 1 & \rho_0 & \rho_4 & \rho_2 &\rho_6   \cr
		s_{211} & \rho_6 & \rho_2 & \rho_4 & \rho_0 & 1 & \rho_3 & \rho_5 &\rho_1   \cr
        s_{212} & \rho_2 & \rho_6 & \rho_0 & \rho_4 & \rho_3 & 1 & \rho_1 &\rho_5   \cr 
        s_{221} & \rho_4 & \rho_0 & \rho_6 & \rho_2 & \rho_5 & \rho_1 & 1 &\rho_3   \cr
		s_{222} & \rho_0 & \rho_4 & \rho_2 & \rho_6 & \rho_1 & \rho_5 & \rho_3 &1   \cr}
\end{align*}
where $s_{k_1k_2k_3}=s(max(|x_{i1k_1}-x_{i2k_2}|,|x_{i1k_1}-x_{i3k_3}|,|x_{i2k_2}-x_{i3k_3}|),\theta)$. For each row, the sum of all elements is equal to $\sum_{l=0}^{7}\rho_l$. 

For the general case with $J$ rater and $n$ replicates,  the $m^{th}$ row sum of the correlation matrix can be expressed as 
\begin{align}
\sum_{m'=1}^M  cor \left(s(D_{im},\theta),s(D_{im'},\theta)\right)
\end{align} 
The correlation between any two $s(D_{im},\theta)$ and $s(D_{im'},\theta)$ depends on the distinct combination of common measurements of $x_{ijk}$ used in $D_{im}$ and $D_{im'}$. As seen in the special case, there are $\begin{pmatrix}J\\l\end{pmatrix}$ combinations where $s(D_{im},\theta)$ and $s(D_{im'},\theta)$ share exactly $h$ common reads of $x_{ijk}$ from $h$ raters out of the $J$ raters for given $m, m'$. Thus there are a total of  $\begin{pmatrix}J\\0\end{pmatrix}+\begin{pmatrix}J\\1\end{pmatrix}+\begin{pmatrix}J\\2\end{pmatrix}+...+\begin{pmatrix}J\\J\end{pmatrix}=2^J$ different such combinations and we denote the corresponding correlations as $\rho_l$, $l=0,\cdots,2^J-1$, i.e., 
\begin{align}
COR\left(s(D_{im},\theta),s(D_{im'},
\theta)\right)=\rho_{l}, \quad l=\sum_{j\in \mathcal{O}_{mm'}} 2^{j-1} \text{ when } \mathcal{O}_{mm'}\neq\emptyset \text{ and } l=0 \text{ when } \mathcal{O}_{mm'}=\emptyset
\end{align}

where $\mathcal{O}_{mm'}$ is the collection of index of the raters whose replicates are common in the two distances. Two extreme correlations are the one when $D_{im}$ and $D_{im'}$ are defined over distinct replicates from each raters, $\mathcal{O}_{mm'}=\emptyset$ and the one when two PMD are defined the same set of the replicates ,$\mathcal{O}_{mm'}=\{1,\cdots,J\}$, which has correlation of 1. For a given $D_{im}$, if $D_{im}$ and $D_{im'}$ contain exactly, say $h$ common reads from $h$ observers, $j_1, j_2, ..., j_h$, then there are $(n-1)^{J-h}$ different $D_{im'}$s such that $COR\left(s(D_{im},\theta),s(D_{im'},\theta)\right)$ is the same to be denoted as $\rho_{2^{j_1-1}+2^{j_2-1}+ ... + 2^{j_h-1}}$. These different $D_{im'}$s come from the $(n-1)^{J-h}$ possible combinations of replications that are different from the ones in $D_{im}$ in the remaining $J-h$ observers. Therefore, the row sum of the $R_0$ could be expressed as
\begin{align}
\small
    a=&(n-1)^J\rho_{0}+\sum_{j=1}^J(n-1)^{J-1}\rho_{2^{j-1}}+\sum_{j_1<j_2}(n-1)^{J-2}\rho_{2^{j_1-1}+2^{j_2-1}}\\
    &+\cdots+\sum_{j_1<\cdots<j_{J-1}}
    (n-1)\rho_{\sum_{t=1}^{J-1}2^{j_t-1}}+1
    \label{equ::rowall}
\end{align}
Since the expression for each row is the same,  Lemma \ref{lemma::euqalrow} is proved.\\

\section{Proof of Theory \ref{th::eq0}\label{app:geeq}}
Under the model assumption discussed at the beginning of Section 2.4 that all $D_{im}$ follow the same marginal distribution, then $s(D_{im},\theta)$ follow same marginal distribution. Let $\sigma_S^2=var(s(D_{im},\theta))$. Then $A_i=diag(cov(S_i))=\mathbf{I}\sigma_S^2$ and $V_i=A_i^{1/2}R_wA_i^{1/2}=\sigma_S^2R_w$. The true covariance matrix is
\begin{align}
cov(S_i)&=\mathbf{A}_i^{1/2}\mathbf{R}_0\mathbf{A}_i^{1/2}=\sigma_S^2 \mathbf{R}_0
\end{align}
where $R_0$ is the true correlation matrix.\\
Recall that $\Gamma_i=\mathbf{1}_{M\times 1}h(\theta)$. Following the general result of GEE, the robust sandwich estimator of  $\hat{\theta}$ under working correlation matrix $R_w$ after plugging in $\mathbf{A}_i=\mathbf{I}\sigma_S^2$ is
 \begin{align}
 V_\theta^{sw} &=N(\Sigma_1)^{-1}\Sigma_0 (\Sigma_1)^{-1} \\
 \Sigma_1&=\sum_{i=1}^N \Gamma_i^T V_i^{-1}\Gamma_i\\
 &=h(\theta)^2\sigma_S^{-2}\sum_{i=1}^N \mathbf{1}_{M\times1}^TR_w^{-1}\mathbf{1}_{M\times1}\\
 &=Nh(\theta)^2 \sigma_S^{-2}\mathbf{1}_{M\times 1}^{T} R_w^{-1}\mathbf{1}_{M\times 1}\\
 \Sigma_0 &=\sum_{i=1}^N  \Gamma_i^T V_i^{-1}cov(S_i)V_i^{-1}\Gamma_i\\
 &=h(\theta)^2\sigma_S^{-2}\sum_{i=1}^N \mathbf{1}_{M\times1}^T R_w^{-1} cov(S_i) R_w^{-1} \mathbf{1}_{M\times1}\\
&=N h(\theta)^2\sigma_S^{-2}\mathbf{1}_{M\times 1}^{T} R_w^{-1}R_0 R_w^{-1} \mathbf{1}_{M\times 1}
 \end{align}
 Now we prove the theorem by showing that the robust sandwich variance estimate of the estimated parameter $\theta$  based on independent working correlation matrix is the same as the variance of the estimated parameter based on true correlation matrix.
\begin{itemize}
 \item  When $R_w=I$ then 
\begin{align}
 \Sigma_1&=MN\sigma_S^{-2}h(\theta)^2\\
  \Sigma_0 &=N\sigma_S^{-2} h(\theta)^2\mathbf{1}_{M\times 1}^{T}R_0  \mathbf{1}_{M\times 1}\\
\end{align}
Based on Lemma \ref{lemma::euqalrow}, the row sums of $R_0$ are equal to a, then
\begin{align}
    \mathbf{1}_{M\times 1}^{T}R_0  \mathbf{1}_{M\times 1}=aM
\end{align}
Therefore, the robust sandwich estimator under working independent correlation matrix is
 \begin{align}\label{equ::v_wi}
V_\theta^{i} &=N(\Sigma_1)^{-1}\Sigma_0 (\Sigma_1)^{-1} \\
&=\frac{a\sigma_S^2}{Mh^2(\theta)}
\end{align}
\item Suppose $R_w=R_0$. Then $\Sigma_1=\Sigma_0$ and $N^{1/2}(\hat{\theta}-\theta_0)$ has variance as 
 \begin{align}\label{equ::v_eq}
V_\theta^{t} &=N\Sigma_0^{-1}= \frac{\sigma_S^2}{h^2(\theta)} (\mathbf{1}_{M\times 1}^{T} R_0^{-1} \mathbf{1}_{M\times 1})^{-1}
\end{align}
 \end{itemize}
 Let $v=\frac{1}{a}\mathbf{1}_{M\times1}$, then $R_0\cdot v=\mathbf{1}_{M\times 1}$.
\begin{align}
    \mathbf{1}_{M\times 1}^T R_0^{-1}\mathbf{1}_{M\times 1}=\mathbf{1}_{M\times 1}^T R_0^{-1} R_0 v=\mathbf{1}_{M \times 1}^T v=\frac{1}{a}\mathbf{1}_{M\times 1}^T\mathbf{1}_{M\times 1}=\frac{M}{a}
\end{align}
Therefore,  the robust sandwich estimator under true correlation matrix is
\begin{align}
V_\theta^t=\frac{a\sigma^2}{Mh^2(\theta)}
\end{align}
Therefore $V_\theta^t=V_\theta^{i}$ implying that we have the same asymptotic efficiency under the true correlation matrix or the independent working correlation. The asymptomatic distribution of $\hat{\theta}$ for agreement study has the following form
 \begin{align}
    N^{1/2}(\hat{\theta}-\theta) \overset{d}{\longrightarrow}\mathcal{N}
    \left(0, \frac{a\sigma^2_S}{Mh^2(\theta)} \right)
    \label{equ::geeout}
\end{align}
where a is the row sum of $R_0$.

\section{Relationship between Log-normal Distribution and Normal Distribution}\label{app::lognormal}
Suppose $Y_i$ follows a log-normal distribution with mean and covariance structure in equations (\ref{equ::simus1})-(\ref{equ::simus3}). Let $X_i=\log(Y_i)$, then $X_i$ should follow a multivariate normal distribution where its mean $\mu_X$ and covariance $\Sigma_X$ have the following form,
\begin{align}
\mathbf{\mu_X}&= (\mu_{1}' \mathbf{1}_{1\times K},\mu_{2}' \mathbf{1}_{1\times K},\mu_{3}' \mathbf{1}_{1\times K})^T \label{equ::simus11}\\
\Sigma_X&=\begin{pmatrix}
{\sigma_1'}^2 \Sigma_1' & \sigma_1' \sigma_2'\rho_{12}' \mathbf{1}_{K\times K} & \sigma_1' \sigma_3'\rho_{13}' \mathbf{1}_{K\times K} \\
 \sigma_1' \sigma_2' \rho_{12}' \mathbf{1}_{K\times K} & {\sigma_2'}^2 \Sigma_2' & \sigma_2' \sigma_3'\rho_{23}' \mathbf{1}_{K\times K} \\
 \sigma_1' \sigma_3'\rho_{13}' \mathbf{1}_{K\times K} & \sigma_2' \sigma_3'\rho_{23}' \mathbf{1}_{K\times K} & {\sigma_3'}^2 \Sigma_3'
\end{pmatrix}\label{equ::simus12}\\
\Sigma_j' &=\begin{pmatrix}
1 & \rho_j' & \rho_j'\\
\rho_j' & 1 & \rho_j' \\
\rho_j' &\rho_j' &1
\end{pmatrix}  \quad \text{for j=1,2 and 3} \label{equ::simus13}
\end{align}
Based on the moment generating function, we could express the parameters of $Y_i$ in terms of parameters of $X_i$ as \cite{halliwell2015lognormal}
\begin{align}
\begin{cases}
E[Y_{ijk}]=\exp(\mu_j'+{\sigma_j'}^2/2)\\
Var[Y_{ijk}]=E[Y_{ijk}]^2\left[\exp({\sigma_j'}^2)-1\right]\\
Cov(E[Y_{ijk_1}],E[Y_{ijk_2}])=E[Y_{ijk_1}]^2\left[\exp({\sigma_j'}^2\rho_j')-1 \right]\\
Cov(E[Y_{ij_1k_1}],E[Y_{ij_2k_2}])=E[Y_{ij_1k_1}]E[Y_{ij_2k_2}]\left[ \exp(\sigma_{j_1}'\sigma_{j_2}'\rho_{j_1 j_2}' )-1 \right]
\end{cases}
\end{align}

Then by setting the left terms to the parameters for Y,  one can solve the equations to obtain the means, variances and covariance for X in term of the parameters for Y as shown below
\begin{align}
\begin{cases}
\mu_j'=2\log(\mu_j)-0.5\log(\sigma_j^2+\mu_j^2)\\
{\sigma_j'}^2=\log(\sigma_j^2+\mu_j^2)-2\log(\mu_j)\\
\rho_j=\frac{\log(\sigma_j^2\rho_j+\mu_j^2)-2\log(\mu_j)}{\log(\sigma_j^2+\mu_j^2)-2\log(\mu_j)}\\
\rho_{j_1j_2}'=\frac{\log(\sigma_{j_2}\sigma_{j_1}\rho_{j_1j_2}+\mu_{j_1}\mu_{j_2})-\log(\mu_{j_1}\mu_{j_2})}{\sqrt{\log(\sigma_{j_1}^2+\mu_{j_1}^2)-2\log(\mu_{j_1})}\sqrt{\log(\sigma_{j_2}^2+\mu_{j_2}^2)-2\log(\mu_{j_2})}}
\end{cases}
\end{align}
We first generate $X_i$ from a multivariate normal distribution with mean and covariate structure from (\ref{equ::simus11})-(\ref{equ::simus13}). Then we set $Y_i=\exp(X_i)$ to obtain the generated log-normal data.
\section{Performance of the new non-parametric estimator for RAUOCPC}\label{app::simu_newraucpc}
We preformed simulation studies to evaluate the performance of this new non-parametric estimator of RAUCPC with comparison to the two exiting estimators. The simulation setting are the same as in Barnhart's paper\cite{raucpc}. Both normally distributed and non-normally distributed data were generated. For normally distributed data, $D_{ijj'}$ were generated from standard normal distribution, while the non-normally distributed $D_{ijj'}$ were generated from a mixture of standard normal and log-normal distribution of $N(0,1)+\mathrm{log}N(0.1, 2.15)$. For both distributions, a total of 1000 simulated data sets were generated with sample sizes of 10, 20, 50, 100 and 200. The $\delta_{max}$ is set to be 2 or 3. The performance of different estimators is evaluated by: (1) bias=estimated RAUCPC-true RAUCPC; (2) MSE=var(estimated RAUCPC)+$\text{bias}^2$; (3) coverage rate: percentage of $95\%$ confident intervals (estimated RAUCPC $\pm$ 1.96se(estimated RAUCPC)) containing true RAUCPC out of 1000 simulations. 

The results are displayed in tables (\ref{tab::simout1}) and (\ref{tab::simout2}). For normally distributed data, the performance of the new proposed non-parametric estimator is similar with the parametric estimator which is better than the bootstrap estimator. When the data is non-normal, the parametric estimator based on normal distribution performed badly as expected where we observe that the 95\% coverage rate decreases to 0 as the sample size increases. While the new proposed estimator have slightly better performance than the bootstrap estimator in terms of the bias and the 95\% coverage.

\begin{sidewaystable}[htbp] 
\centering
\caption{Estimated RAUCPC for Normal Data}\label{tab::simout1}
\begin{tabular}{lllllllllllllll}
\hline
\multirow{2}{*}{$\delta_{max}$} & \multirow{2}{*}{\begin{tabular}[c]{@{}l@{}}Sample\\ Size\end{tabular}} & \multirow{2}{*}{\begin{tabular}[c]{@{}l@{}}True\\ RAUCPC\end{tabular}} & \multicolumn{4}{l}{Parametric Estimator}         & \multicolumn{4}{l}{Non-parametric Bootstrap Estimator} & \multicolumn{4}{l}{New Non-parametric Estimator} \\ \cline{4-15} 
                                &                                                                        &                                                                        & Estimate & Bias   & MSE   & CR & Estimate    & Bias     & MSE     & CR  & Estimate  & Bias  & MSE   & CR \\ \hline
2          & 10           & 0.610        & 0.606               & -0.003          & 0.006          & 0.865                    & 0.648                  & 0.038              & 0.009             & 0.825                       & 0.609        & -0.001   & 0.008   & 0.891                       \\
2          & 20           & 0.610        & 0.611               & 0.001           & 0.003          & 0.901                    & 0.633                  & 0.023              & 0.004             & 0.879                       & 0.612        & 0.002    & 0.004   & 0.931                       \\
2          & 50           & 0.610        & 0.608               & -0.002          & 0.001          & 0.938                    & 0.618                  & 0.008              & 0.001             & 0.942                       & 0.608        & -0.001   & 0.001   & 0.953                       \\
2          & 100          & 0.610        & 0.609               & -0.001          & 0.001          & 0.934                    & 0.614                  & 0.004              & 0.001             & 0.947                       & 0.609        & -0.001   & 0.001   & 0.948                       \\
2          & 200          & 0.610        & 0.610               & 0.000           & 0.000          & 0.945                    & 0.612                  & 0.003              & 0.000             & 0.955                       & 0.610        & 0.000    & 0.000   & 0.953                       \\
3          & 10           & 0.734        & 0.730               & -0.004          & 0.003          & 0.874                    & 0.767                  & 0.032              & 0.004             & 0.802                       & 0.736        & 0.002    & 0.004   & 0.887                       \\
3          & 20           & 0.734        & 0.736               & 0.002           & 0.002          & 0.889                    & 0.756                  & 0.022              & 0.002             & 0.837                       & 0.739        & 0.004    & 0.002   & 0.904                       \\
3          & 50           & 0.734        & 0.733               & -0.001          & 0.001          & 0.924                    & 0.742                  & 0.008              & 0.001             & 0.910                       & 0.734        & 0.000    & 0.001   & 0.929                       \\
3          & 100          & 0.734        & 0.734               & 0.000           & 0.000          & 0.945                    & 0.739                  & 0.004              & 0.000             & 0.934                       & 0.734        & 0.000    & 0.000   & 0.945                       \\
3          & 200          & 0.734        & 0.734               & 0.000           & 0.000          & 0.947                    & 0.737                  & 0.002              & 0.000             & 0.949                       & 0.734        & 0.000    & 0.000   & 0.956           \\
 \hline        
\end{tabular}

\caption{Estimated RAUCPC for Non-normal Data} \label{tab::simout2}
\begin{tabular}{lllllllllllllll}
\hline
\multirow{2}{*}{$\delta_{max}$} & \multirow{2}{*}{\begin{tabular}[c]{@{}l@{}}Sample\\ Size\end{tabular}} & \multirow{2}{*}{\begin{tabular}[c]{@{}l@{}}True\\ RAUCPC\end{tabular}} & \multicolumn{4}{l}{Parametric Estimator}         & \multicolumn{4}{l}{Non-parametric Bootstrap Estimator} & \multicolumn{4}{l}{New Non-parametric Estimator} \\ \cline{4-15} 
                                &                                                                        &                                                                        & Estimate & Bias   & MSE   & CR & Estimate    & Bias     & MSE     & CR  & Estimate  & Bias  & MSE   & CR \\ \hline
2                               & 10                                                                     & 0.442                                                                  & 0.237    & -0.204 & 0.07  & 0.583                & 0.483       & 0.041    & 0.014   & 0.892                  & 0.445     & 0.003 & 0.013 & 0.896                \\
2                               & 20                                                                     & 0.442                                                                  & 0.181    & -0.261 & 0.085 & 0.328                & 0.463       & 0.021    & 0.007   & 0.915                  & 0.442     & 0     & 0.006 & 0.928                \\
2                               & 50                                                                     & 0.442                                                                  & 0.121    & -0.321 & 0.11  & 0.033                & 0.451       & 0.01     & 0.002   & 0.936                  & 0.442     & 0     & 0.002 & 0.943                \\
2                               & 100                                                                    & 0.442                                                                  & 0.09     & -0.352 & 0.127 & 0.001                & 0.447       & 0.005    & 0.001   & 0.937                  & 0.442     & 0     & 0.001 & 0.945                \\
2                               & 200                                                                    & 0.442                                                                  & 0.07     & -0.372 & 0.14  & 0                    & 0.445       & 0.003    & 0.001   & 0.952                  & 0.442     & 0.001 & 0.001 & 0.953                \\
3                               & 10                                                                     & 0.556                                                                  & 0.34     & -0.217 & 0.092 & 0.583                & 0.595       & 0.039    & 0.012   & 0.862                  & 0.559     & 0.003 & 0.011 & 0.892                \\
3                               & 20                                                                     & 0.556                                                                  & 0.25     & -0.306 & 0.123 & 0.388                & 0.579       & 0.023    & 0.006   & 0.901                  & 0.559     & 0.003 & 0.006 & 0.926                \\
3                               & 50                                                                     & 0.556                                                                  & 0.174    & -0.382 & 0.159 & 0.077                & 0.566       & 0.009    & 0.002   & 0.932                  & 0.557     & 0     & 0.002 & 0.941                \\
3                               & 100                                                                    & 0.556                                                                  & 0.133    & -0.423 & 0.186 & 0.001                & 0.562       & 0.006    & 0.001   & 0.939                  & 0.557     & 0.001 & 0.001 & 0.943                \\
3                               & 200                                                                    & 0.556                                                                  & 0.108    & -0.448 & 0.204 & 0                    & 0.56        & 0.003    & 0.001   & 0.949                  & 0.557     & 0.001 & 0.001 & 0.951        \\
 \hline        
\end{tabular}

\end{sidewaystable}
\end{appendices}

\end{document}